\documentclass[twocolumn,preprintnumbers,prd,superscriptaddress,nofootinbib]{revtex4-2}

\usepackage{graphicx,amsmath}
\usepackage{dcolumn}
\usepackage{bm}
\usepackage[dvipsnames,table]{xcolor}
\usepackage{multirow}
\usepackage{verbatim}
\usepackage[T1]{fontenc}
\usepackage[utf8]{inputenc}
\usepackage{booktabs}
\usepackage{transparent}
\usepackage{pifont}
\usepackage{enumitem}
\usepackage{soul}

\usepackage{microtype}
\usepackage{hyperref}
\hypersetup{
  colorlinks=true,
  linkcolor=blue,
  citecolor=blue,
  urlcolor=blue
}
\usepackage[capitalise]{cleveref}

\begin{document}

\title{\texorpdfstring{%
Probing dynamical dark energy with late-time data:\\
Evidence, tensions, and the limits of the $w_0w_a$CDM framework%
}{%
Probing dynamical dark energy with late-time data: Evidence, tensions, and the limits of the w0waCDM framework%
}}

\author{Tengpeng Xu}
\email{tpxu@nao.cas.cn}
\affiliation{School of Physics and Astronomy, Sun Yat-Sen University, Zhuhai 519082, China}
\affiliation{National Astronomical Observatories, Chinese Academy of Sciences Beijing 100101, China}

\author{Suresh Kumar}
\email{suresh.kumar@plaksha.edu.in}
\affiliation{Plaksha University, Mohali, Punjab-140306, India}

\author{Yun Chen}
\email{chenyun@bao.ac.cn}
\affiliation{National Astronomical Observatories, Chinese Academy of Sciences Beijing 100101, China}
\affiliation{College of Astronomy and Space Sciences, University of Chinese Academy of Sciences, Beijing 100049, China}

\author{Abraão J. S. Capistrano}
\email{capistrano@ufpr.br}
\affiliation{Graduate Program in Applied Physics, Federal University of Latin-American Integration, Avenida Tarquínio Joslin dos Santos, 1000 - Polo Universitário, Foz do Iguaçu/PR, Brazil}
\affiliation{Departamento de Engenharias e Ciências Exatas, Universidade Federal do Paraná, Rua Pioneiro, 2153, Palotina/PR, Brazil}

\author{\"{O}zg\"{u}r Akarsu}
\email{akarsuo@itu.edu.tr}
\affiliation{Department of Physics, Istanbul Technical University, Maslak 34469 Istanbul, T\"{u}rkiye}

\begin{abstract}
We test the dynamical dark-energy $w_0w_a$CDM (CPL) framework against $\Lambda$CDM using CMB anisotropies and lensing together with DESI DR2, SDSS-IV, transverse/angular BAO (BAOtr), and Cepheid-calibrated PantheonPlus SN~Ia data. CPL inferences are strongly dataset-dependent. CMB data alone leave a broad geometric degeneracy, while DESI DR2 BAO pulls the reconstruction toward weak present-day acceleration. In contrast, CMB combined with PP\&SH0ES and BAOtr favors a moderately accelerating expansion and substantially reduces the Hubble tension. The origin of this behavior can be traced to low-redshift distance information: BAOtr and DESI/SDSS prefer different BAO distance ratios at $z\lesssim0.5$, which drives divergent CPL expansion histories. We quantify this mismatch directly at the data level by comparing angular BAO scales, including the common $z=0.510$ point and a conservative local interpolation of BAOtr with no extrapolation. As expected within CPL, where pre-recombination physics is fixed, $r_{\rm d}$ remains nearly unchanged, so shifts in $H_0$ are absorbed by late-time expansion freedom rather than by a change in the sound horizon. Bayesian evidence is likewise contingent on the low-redshift data: it favors CPL mainly when PP\&SH0ES and/or BAOtr are included, is inconclusive for CMB-only and CMB+DESI, and moderately favors $\Lambda$CDM for CMB+SDSS. These results show that apparent support for CPL and its ability to ease the Hubble tension are not universal, motivating more flexible late-time models and closer scrutiny of BAO systematics.
\end{abstract}

\maketitle

\section{Introduction}
\label{sec:intro}

Late-time cosmic acceleration is supported by a broad and mutually reinforcing set of observations, spanning geometric probes of the background (Type~Ia supernovae and baryon acoustic oscillations, BAO) and the growth of structure (weak lensing and clustering), all anchored by high-precision measurements of the cosmic microwave background (CMB)
\cite{Planck:2018vyg,Planck:2018nkj,AtacamaCosmologyTelescope:2025blo,SPT-3G:2025bzu,eBOSS:2020yzd,DESI:2025zgx,Scolnic:2021amr,Brout:2022vxf,Rubin:2023jdq,DES:2025sig}.
Within general relativity (GR), the minimal phenomenological description is a spatially flat $\Lambda$ cold dark matter model ($\Lambda$CDM), in which acceleration is driven by a strictly constant vacuum-energy density with equation of state $w=-1$.
Despite its empirical success, the physical origin of this component remains unknown, and the cosmological-constant problem continues to motivate precision tests of whether late-time data require a rigid $\Lambda$ or permit (or even prefer) late-time dynamics in an effective dark sector~\cite{Weinberg:1988cp,Sahni:2002kh}.

A second motivation is internal consistency.
As cosmological measurements have entered the percent era, multiple parameters inferred from distinct probes exhibit mild-to-moderate discordances that may signal residual systematics, underestimated covariances, or limitations of the minimal model
\cite{Verde:2019ivm,DiValentino:2020zio,DiValentino:2021izs,Perivolaropoulos:2021jda,Abdalla:2022yfr,DiValentino:2022fjm,Khalife:2023qbu,Vagnozzi:2023nrq,Akarsu:2024qiq,CosmoVerseNetwork:2025alb}.
The best-known example is the \emph{Hubble tension}~\cite{Verde:2019ivm,DiValentino:2020zio,DiValentino:2021izs}, the mismatch between local determinations of the Hubble constant and values inferred from early-universe data analyzed within $\Lambda$CDM.
Cepheid-calibrated SN~Ia measurements from SH0ES give $H_0\simeq 73~{\rm km\,s^{-1}\,Mpc^{-1}}$~\cite{Breuval:2024lsv,Riess:2021jrx}, while CMB-based analyses prefer significantly smaller values~\cite{Planck:2018vyg,SPT-3G:2025bzu}. Using the recent CMB-inferred value $H_0=67.24\pm0.35~{\rm km\,s^{-1}\,Mpc^{-1}}$ from Planck+SPT+ACT~\cite{SPT-3G:2025bzu} and the $H_0$ Distance Network (H0DN; the Local Distance Network compilation) value ${H_0 = 73.50 \pm 0.81~{\rm km\,s^{-1}\,Mpc^{-1}}}$~\cite{H0DN:2025lyy}
as a representative comparison yields a discrepancy at the $\simeq 7.1\sigma$ level. In this work we adopt the H0DN compilation as the local reference because it provides a covariance-weighted, community-vetted synthesis of several leading local distance indicators, thereby reducing dependence on any single calibration route~\cite{H0DN:2025lyy}.

Proposed extensions that address $H_0$ and related late-time discrepancies are often classified according to when they modify the expansion history~\cite{CosmoVerseNetwork:2025alb}.
\emph{Early-time} solutions change pre-recombination physics and typically shift the sound-horizon scale at the baryon-drag epoch, $r_{\rm d}$, e.g.\ early dark energy (EDE)~\cite{Poulin:2018cxd,Karwal:2016vyq,Hill:2020osr,Kamionkowski:2022pkx,Ivanov:2020ril,Sakstein:2019fmf,Niedermann:2019olb,Niedermann:2020dwg,Poulin:2023lkg,Smith:2025grk,Poulin:2025nfb,SPT-3G:2025vyw}.
By contrast, \emph{late-time} solutions modify the post-recombination expansion history while leaving $r_{\rm d}$ essentially unchanged when pre-recombination physics is standard, e.g.\ interacting dark energy (IDE)~\cite{Caprini:2016qxs,Nunes:2016dlj,Kumar:2017dnp,DiValentino:2017iww,Yang:2017ccc,Costa:2018aoy,vonMarttens:2018iav,Yang:2018euj,Yang:2018uae,Pan:2019gop,Kumar:2019wfs,DiValentino:2019jae,DiValentino:2019ffd,DiValentino:2020kpf,Gomez-Valent:2020mqn,Lucca:2020zjb,Pan:2020zza,Gao:2021xnk,Kumar:2021eev,Yang:2021hxg,Nunes:2022bhn,Bernui:2023byc,Escamilla:2023shf,Giare:2024smz,Li:2024qso,Li:2025owk,Sabogal:2025mkp,Silva:2025hxw,Yang:2025uyv,vanderWesthuizen:2025rip,Li:2026xaz} and dynamical dark-energy scenarios exhibiting AdS-to-dS(-like) transitions in the late Universe (at redshift $\sim2$), such as $\Lambda_{\rm s}$CDM~\cite{Akarsu:2019hmw,Akarsu:2021fol,Akarsu:2022typ,Akarsu:2023mfb,Paraskevas:2024ytz,Yadav:2024duq,Akarsu:2024qsi,Akarsu:2024eoo,Akarsu:2024nas,Souza:2024qwd,Akarsu:2025gwi,Akarsu:2025dmj,Akarsu:2025ijk,Escamilla:2025imi,Akarsu:2025nns,Kibris:2026cqq} (see also~\cite{Anchordoqui:2023woo,Anchordoqui:2024gfa,Anchordoqui:2024dqc,Soriano:2025gxd}), which present particularly economical extensions of standard $\Lambda$CDM, as well as more complicated constructions, e.g.\ omnipotent dark energy (allowing rich dynamics such as phantom crossing and sign-changing energy density in the late Universe)~\cite{DiValentino:2020naf,Adil:2023exv,Specogna:2025guo}. For further reading on related theoretical and observational studies and model-agnostic reconstructions, see Refs.~\cite{Sahni:2002dx,Vazquez:2012ag,BOSS:2014hwf,Sahni:2014ooa,BOSS:2014hhw,DiValentino:2017rcr,Mortsell:2018mfj,Poulin:2018zxs,Wang:2018fng,Banihashemi:2018oxo,Dutta:2018vmq,Banihashemi:2018has,Li:2019yem,Akarsu:2019ygx,Visinelli:2019qqu,Perez:2020cwa,Akarsu:2020yqa,Ruchika:2020avj,Calderon:2020hoc,DeFelice:2020cpt,Paliathanasis:2020sfe,Bonilla:2020wbn,Acquaviva:2021jov,Bag:2021cqm,Bernardo:2021cxi,Escamilla:2021uoj,Sen:2021wld,Ozulker:2022slu,DiGennaro:2022ykp,Akarsu:2022lhx,Moshafi:2022mva,vandeVenn:2022gvl,Ong:2022wrs,Tiwari:2023jle,Malekjani:2023ple,Vazquez:2023kyx,Alexandre:2023nmh,Adil:2023ara,Paraskevas:2023itu,Gomez-Valent:2023uof,Wen:2023wes,Wen:2024orc,DeFelice:2023bwq,Menci:2024rbq,Gomez-Valent:2024tdb,DESI:2024aqx,Bousis:2024rnb,Wang:2024hwd,Colgain:2024ksa,Tyagi:2024cqp,Toda:2024ncp,Sabogal:2024qxs,Dwivedi:2024okk,Escamilla:2024ahl,Gomez-Valent:2024ejh,Manoharan:2024thb,Pai:2024ydi,Mukherjee:2025myk,Efstratiou:2025xou,Gomez-Valent:2025mfl,Wang:2025dtk,Bouhmadi-Lopez:2025ggl,Tamayo:2025xci,Gonzalez-Fuentes:2025lei,Bouhmadi-Lopez:2025spo,Hogas:2025ahb,Yadav:2025vpx,Lehnert:2025izp,Tan:2025xas,Pedrotti:2025ccw,Forconi:2025gwo,Nyergesy:2025lyi,Ghafari:2025eql,Akarsu:2026anp,Gokcen:2026pkq,Akarsu:2026pom}.
Along related lines, model-independent reconstructions of IDE kernels do not rule out negative effective DE densities at $z\gtrsim 2$~\cite{Escamilla:2023shf}.
Late-time modifications are primarily tested by the internal consistency of low-redshift distance information (e.g.\ SN~Ia and BAO) once early-universe data (e.g.\ the CMB) calibrate the relevant physical scale (most notably through $r_{\rm d}$).
Robustness to the choice of low-$z$ distance data is therefore a key issue for phenomenological late-time modeling.

In particular, a growing body of work shows that extensions such as \textit{dynamical dark energy} (DDE) can substantially improve the joint consistency of BAO and supernova data relative to $\Lambda$CDM.
Recent BAO measurements from DESI~\cite{DESI:2024mwx,DESI:2025zgx} provide high-precision distance ratios over a wide redshift range and have prompted renewed scrutiny of late-time model extensions, particularly DDE scenarios~\cite{Giare:2024gpk,Gialamas:2024lyw,Gialamas:2025qph,RoyChoudhury:2024wri,Dinda:2024kjf,Giare:2024oil,RoyChoudhury:2025dhe,RoyChoudhury:2025iis,Scherer:2025esj,Pang:2025lvh,Roy:2024kni,Ormondroyd:2025iaf,Li:2025cxn,Li:2025vuh,Cortes:2024lgw,Najafi:2024qzm,Wang:2024dka,Giare:2025pzu,Kessler:2025kju,Teixeira:2025czm,Sabogal:2025jbo,Cheng:2025lod,Herold:2025hkb,Cheng:2025hug,Ozulker:2025ehg,Lee:2025pzo,Silva:2025twg,Fazzari:2025lzd,Bouhmadi-Lopez:2025lzm,Mishra:2025goj,Wolf:2024set,Wolf:2025ace,Wolf:2025mco,Wolf:2025rde,Wolf:2026ccg}, including their potential relation to the $H_0$ tension.
In the DESI DR2 analysis, combinations including CMB and BAO have been reported to prefer an evolving dark-energy equation of state within common parameterizations for some dataset choices, while other studies emphasize the dependence on priors, degeneracies, and the specific selection of low-redshift data~\cite{DESI:2025zgx,Giare:2025pzu}.
A recent multi-model DESI DR2 analysis combining CMB, BAO, and SN~Ia data similarly finds evidence for dynamical dark energy in several late-time frameworks, while the inferred $H_0$ remains close to early-universe values and hence leaves the Hubble tension unresolved~\cite{Zhang:2026dde}.
This motivates a concrete question that can be addressed with current data: \emph{are the inferred late-time dynamics stable under reasonable changes in the low-redshift distance dataset, and when an evolving-DE preference is found, does it correlate with alleviating (or worsening) the $H_0$ tension?}

A widely used phenomenological description of dynamical dark energy is the Chevallier--Polarski--Linder (CPL) equation-of-state parameterization~\cite{Chevallier:2000qy,Linder:2002et},
\begin{equation}
w(a)=w_0+w_a(1-a),
\end{equation}
which may be viewed as a first-order expansion about $a=1$ and is therefore most directly interpretable at low redshift.
In this framework the high-redshift asymptote is $w(z\to\infty)=w_0+w_a$, and the corresponding dark-energy density evolves as
$\rho_{\rm de}(z)=\rho_{{\rm de},0}(1+z)^{3(1+w_0+w_a)}\exp\!\left[-3w_a\,\frac{z}{1+z}\right]$, so that for sufficiently large $z$ one has $\rho_{\rm de}\propto (1+z)^{3(1+w_0+w_a)}$ up to an overall constant factor. Because CPL is compact, it is convenient for comparing analyses, but constraints can align along degeneracy directions that correspond to qualitatively different background histories when the model is extrapolated beyond the redshift range directly anchored by data.
This behavior has been discussed in the context of DESI-era fits, including the mapping of degeneracy bands to phantom-crossing behavior~\cite{Giare:2025pzu,Ozulker:2025ehg}.
Relatedly, several alternative two-parameter ans\"atze (e.g.\ JBP, BA, logarithmic, exponential) can yield qualitatively similar ``evolving-DE'' preferences for certain dataset choices~\cite{Jassal:2005qc,Barboza:2008rh,Efstathiou:1999tm,Dimakis:2023oje,Giare:2024gpk}, and effective-fluid descriptions can become ill-defined if an effective $\rho_{\rm de}$ approaches zero and changes sign, since $w_{\rm de}=p_{\rm de}/\rho_{\rm de}$ then ceases to be a stable diagnostic~\cite{Giare:2024gpk}.
For these reasons, CPL is best treated as a diagnostic framework whose utility can be assessed by its dataset-level consistency and by model comparison, rather than by fit improvement alone.

In this paper we test $\Lambda$CDM against $w_0w_a$CDM (CPL) with a focus on robustness to low-redshift dataset choice and on internal consistency among late-time distance probes.
We combine CMB anisotropies and lensing with: (i) standard three-dimensional BAO distances from DESI DR2~\cite{DESI:2025zgx} and from the completed SDSS-IV (BOSS+eBOSS) BAO consensus~\cite{eBOSS:2020yzd} compilation, (ii) an angular/transverse BAO compilation (BAOtr)~\cite{Nunes:2020hzy}, and (iii) the Cepheid-calibrated PantheonPlus SN~Ia likelihood (PP\&SH0ES)~\cite{Scolnic:2021amr,Brout:2022vxf,Riess:2021jrx}.
Beyond constraints in the $(w_0,w_a)$ plane, we reconstruct the implied background history via the conformal Hubble rate $H(z)/(1+z)$ and the deceleration parameter $q(z)$ and discuss their present-day values, $H_0$ and $q_0$, and we track $r_{\rm d}$ and the combination $r_{\rm d}H_0$ to diagnose, within the fixed pre-recombination physics assumed in these models, whether shifts in $H_0$ are absorbed by late-time expansion freedom rather than by changes in the sound-horizon calibration.
We also perform Bayesian model comparison between $\Lambda$CDM and CPL to assess whether any preference for additional late-time parameters is supported once parameter volume is accounted for.

The remainder of the paper is organized as follows.
In Sec.~\ref{sec:data} we describe the datasets and inference strategy.
In Sec.~\ref{sec:results_discussion} we present posterior constraints and reconstructed expansion histories in $\Lambda$CDM and CPL, quantify parameter-level inconsistencies (including in $H_0$), and examine Bayesian evidence as a function of dataset combination, with particular attention to how low-redshift BAO distance ratios propagate into CPL reconstructions.
We summarize our conclusions in Sec.~\ref{sec:conclusion}.

\section{Data and Methodology}
\label{sec:data}

In this section we summarize the datasets and inference strategy used to constrain $\Lambda$CDM and its dynamical dark-energy extension, $w_0w_a$CDM (CPL). Our aim is to assess (i) how different late-time distance probes shape the inferred dark-energy dynamics and (ii) whether the CPL framework can reduce the Hubble tension without introducing new internal inconsistencies among low-redshift datasets. We combine early-universe information from CMB anisotropies and lensing with late-universe distance measurements from BAO and SN~Ia, including DESI DR2 BAO, the completed SDSS-IV/eBOSS BAO consensus dataset, an angular/transverse BAO compilation (BAOtr), and the Cepheid-calibrated PantheonPlus SN~Ia likelihood (PP\&SH0ES).

\subsection{Cosmological Datasets}
\label{subsec:data}

\subsubsection{\textit{Planck}+ACT CMB}

We use the \textit{Planck} 2018 temperature and polarization power spectra (TT, TE, EE), which provide high-precision measurements of primary CMB anisotropies and serve as the cornerstone of modern cosmological parameter estimation~\cite{Planck:2018vyg}. These spectra are sensitive to a broad range of parameters, including the matter content, baryon density, and the background geometry (and hence the inferred expansion history) of the universe.

In addition to the primary power spectra, we also include the lensing potential power spectrum $C^{\phi\phi}_\ell$, which encodes information about the projected large-scale structure through which CMB photons travel. This signal, derived from the four-point correlation function of the CMB, offers a complementary probe of late-time gravitational potentials and helps to break parameter degeneracies relevant for late-time extensions~\cite{Planck:2013mth,Planck:2015mym,Planck:2018lbu}.

For CMB lensing, we adopt the combination of the \textit{Planck} NPIPE PR4 reconstruction~\cite{Carron:2022eyg} and the DR6 lensing measurements from the Atacama Cosmology Telescope (ACT)~\cite{ACT:2023kun,ACT:2023dou,ACT:2023ubw}, collectively referred to as \textit{Planck}+ACT lensing.

Throughout this work, the term ``CMB data'' refers to the combination of \textit{Planck} 2018 TT, TE, and EE spectra along with \textit{Planck}+ACT lensing. These data jointly constrain both early- and late-universe physics, providing a robust foundation for cosmological inference.

\subsubsection{SN Ia and Cepheids}

For SN Ia data, we adopt the latest Pantheon+ compilation~\cite{Brout:2022vxf}, which includes 1701 light curves spanning the redshift range $z \in [0.001, 2.26]$, corresponding to 1550 unique SN Ia events.

To assess implications for the Hubble tension, we incorporate the Cepheid-based distance calibration from the SH0ES collaboration~\cite{Riess:2021jrx}, applied to the Pantheon+ sample. This is implemented by integrating the SH0ES Cepheid host-galaxy distance anchors into the Pantheon+ SN Ia likelihood, following the methodology outlined in~\cite{Scolnic:2021amr,Brout:2022vxf}. The resulting dataset, which combines Pantheon+ with the SH0ES calibration, is referred to as PantheonPlus\&SH0ES (or PP\&SH0ES for short).

\subsubsection{Angular/transverse BAO}

For BAO data, we adopt a set of angular/transverse BAO measurements, referred to as BAOtr, which provide a complementary distance probe with reduced dependence on the fiducial cosmology assumptions used in standard three-dimensional BAO analyses. The BAOtr dataset, compiled by~\cite{Nunes:2020hzy}, consists of 15 measurements of the angular BAO scale, $\theta_{\rm BAO}(z)$, obtained from various SDSS data releases (DR7, DR10, DR11, DR12)~\cite{Carvalho:2015ica,Alcaniz:2016ryy,Carvalho:2017tuu,deCarvalho:2017xye,eBOSS:2020yzd}. These measurements follow:
\begin{equation}
\theta_{\rm BAO}(z)=\frac{r_{\rm d}}{D_{\rm M}(z)},
\end{equation}
where $r_{\rm d}$ is the sound horizon at the drag epoch, and $D_{\rm M}(z)$ is the comoving angular diameter distance.

Compared to standard three-dimensional BAO constraints, BAOtr generally has weaker constraining power, with precision varying substantially across redshift because the information comes from angular clustering in thin redshift slices. A practical advantage of the BAOtr compilation is that its redshift bins are provided as independent measurements from thin slices; in the absence of a public full covariance matrix, we therefore follow the standard treatment and neglect inter-bin covariances in the BAOtr likelihood.

\subsubsection{DESI DR2 BAO}

The Dark Energy Spectroscopic Instrument (DESI) has recently released its second data release (DR2)~\cite{DESI:2025zgx}, providing the most precise and comprehensive baryon acoustic oscillation (BAO) measurements to date. The dataset covers a wide redshift range from 0.295 to 2.33, using three distinct tracers: the Bright Galaxy Sample (BGS), the luminous red galaxies (LRGs), and the emission line galaxies (ELGs), as well as the Ly$\alpha$ forest from high-redshift quasars.

The DESI DR2 BAO data exhibit unprecedented precision, significantly enhancing the ability to constrain cosmological models, particularly those involving dark energy. In the DESI analysis~\cite{DESI:2025zgx}, combining DESI DR2 BAO with \textit{Planck} 2018 CMB data yields a $3.1\sigma$ preference for evolving dark energy within the CPL framework relative to $\Lambda$CDM. When additional SN Ia samples are included, the preference depends on the adopted SN compilation and ranges from $2.8\sigma$ (Pantheon+) to $4.2\sigma$ (DESY5), with Union3 giving $3.8\sigma$~\cite{DESI:2025zgx}.

Given its high precision and broad redshift coverage, DESI DR2 BAO plays a critical role in testing the dynamical nature of dark energy. In this work, we refer to this dataset simply as ``DESI'' and use it as one of the primary probes for confronting theoretical dark energy models.

\subsubsection{Pre-DESI BAO (completed SDSS-IV/eBOSS consensus) Compilation}

In addition to the DESI DR2 BAO measurements, we adopt as our pre-DESI benchmark the final BAO consensus results from the completed SDSS-IV program (BOSS+eBOSS)~\cite{eBOSS:2020yzd}. 
This dataset, hereafter denoted as \textbf{SDSS}, represents the culmination of more than two decades of spectroscopic galaxy surveys conducted at the Apache Point Observatory and provides the most comprehensive and homogeneous pre-DESI three-dimensional BAO compilation currently available.

The SDSS BAO sample consists of 14 measurements, combining information from the BOSS and eBOSS galaxy samples as well as high-redshift quasar and Ly$\alpha$ forest observations. 
These data constrain combinations of the comoving angular diameter distance $D_{\rm M}(z)$, the Hubble distance $D_{\rm H}(z)$, and the spherically averaged distance $D_{\rm V}(z)$ in units of the sound horizon at the drag epoch $r_{\rm d}$, with the full covariance matrix provided by the SDSS-IV collaboration.

The SDSS BAO compilation effectively constrains the late-time expansion history and provides a natural pre-DESI reference for comparison with DESI DR2 BAO, enabling a clearer assessment of how different BAO datasets impact dynamical dark-energy inferences.

For reproducibility, the BAO measurements used in this work are listed explicitly in Appendix~\ref{app:bao_data}. The tabulated uncertainties are the diagonal $1\sigma$ errors; for correlated DESI and SDSS/eBOSS measurements we use the full covariance matrices or likelihood grids provided by the corresponding public likelihoods.

\subsection{Methodology}
\label{sec:method}

To explore the implications of dynamical dark energy, we perform Bayesian parameter estimation and model comparison using the \texttt{Cobaya} framework~\cite{Torrado:2020dgo,2019ascl.soft10019T} with the nested sampling algorithm implemented in \texttt{PolyChord}~\cite{Handley:2015fda,Handley:2015vkr}. This approach allows us to obtain both the posterior distributions of cosmological parameters and robust estimates of the Bayesian evidence for different dark energy models.

We consider two theoretical frameworks: the standard $\Lambda$CDM model and the CPL parameterization of dynamical dark energy ($w_0w_a$CDM), where the dark energy equation of state evolves as
\begin{equation}
w(a) = w_0 + w_a (1-a),
\end{equation}
equivalently $w(z)=w_0+w_a\,z/(1+z)$. Here, $w_0$ and $w_a$ are free parameters that capture the present-day value and the evolutionary rate of the dark energy equation of state, respectively.

We summarize the sampling parameters and their priors used in \texttt{Cobaya} in Table~\ref{tab:prior}. In addition, we define derived parameters including $\Omega_{\rm m}$, $r_{\rm d}$, and $r_{\rm d} H_0$, which are used for further analysis. We assume the standard minimal neutrino sector used in the baseline CMB likelihood setup, with the summed neutrino mass fixed to $\sum m_\nu=0.06\,{\rm eV}$.

\begin{table}
  \centering
 \caption{\label{tab:prior} Priors for the $\Lambda$CDM and $w_0w_a$CDM models. We follow the prior choices used in the DESI DR2 analysis~\cite{DESI:2025zgx}, including the condition $w_0 + w_a < 0$, which helps to avoid early-time dark energy domination in the CPL parametrization. Unlike the DESI analysis, we sample directly in $H_0$ (with a flat prior) rather than using the $100\theta_{\rm MC}$ proxy, ensuring a more transparent interpretation of parameter shifts driven by low-redshift data.}
  \setlength{\tabcolsep}{5pt}
\renewcommand{\arraystretch}{1}
  \begin{tabular}{c|cc}
    \hline
    parameterization & parameter & prior \\
    \hline
    baseline & $\Omega_\mathrm{b} h^2$ & $\mathcal{U}[0.005, 0.1]$ \\
    ($\Lambda$CDM) & $\Omega_\mathrm{c} h^2$ & $\mathcal{U}[0.001, 0.99]$ \\
    & $H_0$ & $\mathcal{U}[20, 100]$ \\
    & $\log(10^{10} A_\mathrm{s})$ & $\mathcal{U}[1.61, 3.91]$ \\
    & $n_\mathrm{s}$ & $\mathcal{U}[0.8, 1.2]$ \\
    & $\tau_\mathrm{reio}$ & $\mathcal{U}[0.01, 0.8]$ \\
    \hline
    extended & $w_0$ & $\mathcal{U}[-3, 1]$ \\
    ($w_0w_a$CDM) & $w_a$ & $\mathcal{U}[-3, 2]$ \\
    \hline
  \end{tabular}
\end{table}

Our analysis is based on the following combinations of cosmological datasets:
\begin{itemize}
  \item \textbf{CMB}: the \textit{Planck}+ACT datasets described in Section~\ref{subsec:data};
  \item \textbf{CMB+SDSS}: combining CMB data with the final pre-DESI BAO consensus results from the completed SDSS-IV program (BOSS+eBOSS)~\cite{eBOSS:2020yzd}, which comprise 14 BAO measurements providing a robust pre-DESI benchmark for comparison;
  \item \textbf{CMB+DESI}: combining CMB data with the latest DESI DR2 BAO measurements for comparison with contemporary large-scale structure constraints;
  \item \textbf{CMB+BAOtr}: incorporating the two-dimensional (angular/transverse) BAO measurements, which provide a complementary late-time distance probe with reduced dependence on the fiducial cosmology assumptions used in standard three-dimensional BAO analyses, though with weaker constraining power;
  \item \textbf{CMB+PP\&SH0ES}: combining the CMB data with the PantheonPlus sample calibrated by SH0ES SN Ia measurements;
  \item \textbf{CMB+PP\&SH0ES+BAOtr}: extending the previous combination by further including the angular/transverse BAO measurements.
\end{itemize}

We focus on three key sets of parameters: the CPL parameters $(w_0, w_a)$, which directly probe the dynamical nature of dark energy; the Hubble constant $H_0$, whose comparison with a recent high-precision local determination (H0DN) provides a quantitative assessment of the Hubble tension; and the sound horizon $r_{\mathrm d}$ and its combination with the Hubble constant, $r_{\mathrm d} H_0$, which help diagnose how, in a framework with standard pre-recombination physics, discrepancies among data combinations are absorbed by $r_{\mathrm d}H_0$ and the late-time expansion history rather than by an independently varying sound horizon.

For each dataset combination and model, we compute posterior constraints on these parameters using weighted samples from the \texttt{PolyChord} chains. We quantify the tension in $H_0$ by comparing posterior samples from different dataset combinations, avoiding Gaussian assumptions so that the results remain reliable even for skewed or non-Gaussian distributions.

Specifically, given two posterior samples $x_A$ and $x_B$ of a parameter $x$ from independent dataset combinations $A$ and $B$, we construct the difference distribution $\Delta x = x_A - x_B$. We then define a two-sided probability
\begin{equation}
p = \min \left[\, P(\Delta x > 0),\, P(\Delta x < 0) \,\right],
\end{equation}
where $P(\Delta x > 0)$ and $P(\Delta x < 0)$ are estimated directly from the samples. The corresponding tension in units of Gaussian standard deviations is defined as
\begin{equation}
\label{eqn:T}
T_\sigma = \Phi^{-1}(1-p),
\end{equation}
where $\Phi^{-1}$ is the inverse cumulative distribution function of the standard normal distribution. This definition provides an equivalent ``$\sigma$ tension'' without assuming Gaussianity of the underlying posteriors.
For nearly Gaussian and independent posteriors this sample-based definition reduces to the familiar estimate $|\mu_A-\mu_B|/\sqrt{\sigma_A^2+\sigma_B^2}$; we use the full posterior-sample construction as the default because several CPL constraints are non-Gaussian or one-sided.

\section{Results and Discussion}
\label{sec:results_discussion}

In this section we present constraints on $\Lambda$CDM and $w_0w_a$CDM (CPL) using the six dataset combinations defined in~\cref{sec:data}. We first summarize the posterior constraints in the $(w_0,w_a)$ plane and their correlation with $H_0$, and quantify the resulting $H_0$ tension relative to H0DN (\cref{subsecIIIA};~\cref{fig:w0waH0_contours,fig:h0_tension}). We then reconstruct the implied late-time expansion history through $H(z)$ and the deceleration parameter $q(z)$, highlighting the strong dataset dependence of CPL reconstructions (\cref{subsecIIIB};~\cref{fig:H_q_z,fig:w0q0omegam}). Model comparison using Bayesian evidence is presented in~\cref{subsecIIIC} (\cref{fig:BE}), and we finally diagnose the origin of the dataset dependence by comparing BAO distance ratios, presenting the direct BAOtr--3D BAO data-level diagnostic, and assessing the consistency of individual probes (\cref{subsecIIID};~\cref{fig:d_z,fig:rdH0Omegam,fig:baotr_interpolation_visualization,fig:w0wacdm_w0wa,fig:individual_contour}).

\newcommand{\twomodel}[2]{%
  $\begin{array}{c}
  #1\\[1.1pt]
  \textcolor{blue}{#2}
  \end{array}$}

\newcommand{\twomodelNA}[1]{%
  $\begin{array}{c}
  #1\\[1.1pt]
  \textcolor{blue}{\text{--}}
  \end{array}$}
\begin{table*}[t]
\centering
\caption{\label{tab:parameters}
Marginalized constraints (mean values with 68\% CL) on the baseline and selected derived parameters for $w_0w_a$CDM (CPL) and $\Lambda$CDM from various dataset combinations. In each entry, the \emph{top} value corresponds to $w_0w_a$CDM (black), while the bottom value corresponds to $\Lambda$CDM (blue). The top block lists sampled parameters and the lower blocks report derived parameters and fit statistics. One-sided limits are quoted at 95\% CL. We define $\Delta\chi^2_{\min}\equiv\chi^2_{\min}(\mathrm{CPL})-\chi^2_{\min}(\Lambda\mathrm{CDM})$ and $\Delta\ln\mathcal{Z}\equiv\ln\mathcal{Z}(\mathrm{CPL})-\ln\mathcal{Z}(\Lambda\mathrm{CDM})$.
}
\setlength{\tabcolsep}{4.1pt}      
\renewcommand{\arraystretch}{1.18} 
\footnotesize
\resizebox{\textwidth}{!}{%
\begin{tabular}{lcccccc}
\toprule
 & CMB & CMB+SDSS & CMB+DESI & CMB+BAOtr & CMB+PP\&SH0ES & \shortstack{CMB+PP\&SH0ES\\+BAOtr}\\
\midrule

$\Omega_{\rm b} h^2$
& \twomodel{0.02243\pm0.00014}{0.02237\pm0.00014}
& \twomodel{0.02238\pm0.00014}{0.02243^{+0.00014}_{-0.00013}}
& \twomodel{0.02240\pm0.00013}{0.02256\pm0.00013}
& \twomodel{0.02244\pm0.00014}{0.02261\pm0.00014}
& \twomodel{0.02248\pm0.00014}{0.02262\pm0.00013}
& \twomodel{0.02249\pm0.00015}{0.02275\pm0.00014}\\

$\Omega_{\rm c} h^2$
& \twomodel{0.1193\pm0.0012}{0.1201\pm0.0012}
& \twomodel{0.1202\pm0.0011}{0.11949\pm0.00090}
& \twomodel{0.11969^{+0.00089}_{-0.00081}}{0.11776^{+0.00064}_{-0.00058}}
& \twomodel{0.1193\pm0.0011}{0.1171\pm0.0010}
& \twomodel{0.1190\pm0.0012}{0.11765^{+0.00096}_{-0.0011}}
& \twomodel{0.1188\pm0.0012}{0.11596\pm0.00092}\\

$\log(10^{10}A_s)$
& \twomodel{3.038\pm0.013}{3.046\pm0.013}
& \twomodel{3.042\pm0.014}{3.050\pm0.013}
& \twomodel{3.043\pm0.013}{3.059^{+0.012}_{-0.014}}
& \twomodel{3.039\pm0.013}{3.061^{+0.013}_{-0.015}}
& \twomodel{3.042\pm0.013}{3.061^{+0.013}_{-0.015}}
& \twomodel{3.042\pm0.015}{3.070^{+0.014}_{-0.017}}\\

$n_s$
& \twomodel{0.9671\pm0.0041}{0.9652\pm0.0041}
& \twomodel{0.9652\pm0.0041}{0.9670\pm0.0037}
& \twomodel{0.9664\pm0.0036}{0.9715\pm0.0033}
& \twomodel{0.9674\pm0.0040}{0.9730\pm0.0039}
& \twomodel{0.9682\pm0.0041}{0.9720\pm0.0039}
& \twomodel{0.9684\pm0.0044}{0.9764\pm0.0038}\\

$\tau_{\rm reio}$
& \twomodel{0.0526\pm0.0074}{0.0547\pm0.0073}
& \twomodel{0.0530\pm0.0075}{0.0570^{+0.0068}_{-0.0077}}
& \twomodel{0.0539\pm0.0069}{0.0621^{+0.0067}_{-0.0079}}
& \twomodel{0.0528\pm0.0074}{0.0634^{+0.0072}_{-0.0085}}
& \twomodel{0.0546\pm0.0072}{0.0632^{+0.0072}_{-0.0083}}
& \twomodel{0.0548\pm0.0084}{0.0692^{+0.0077}_{-0.0093}}\\

$H_0\,\rm [km/s/Mpc]$
& \twomodel{>73.9}{67.31^{+0.49}_{-0.56}}
& \twomodel{63.6^{+2.2}_{-2.5}}{67.59\pm0.41}
& \twomodel{63.9\pm2.0}{68.40\pm0.28}
& \twomodel{73.4^{+2.2}_{-3.8}}{68.69\pm0.48}
& \twomodel{70.87\pm0.68}{68.50\pm0.46}
& \twomodel{71.31\pm0.67}{69.29\pm0.43}\\

$w_0$
& \twomodelNA{-1.40^{+0.33}_{-0.40}}
& \twomodelNA{-0.48\pm0.25}
& \twomodelNA{-0.45^{+0.20}_{-0.23}}
& \twomodelNA{-0.80^{+0.39}_{-0.18}}
& \twomodelNA{-0.694\pm0.078}
& \twomodelNA{-0.660\pm0.079}\\

$w_a$
& \twomodelNA{<1.03}
& \twomodelNA{-1.51^{+0.75}_{-0.68}}
& \twomodelNA{-1.65^{+0.63}_{-0.55}}
& \twomodelNA{<0.132}
& \twomodelNA{-1.70^{+0.40}_{-0.35}}
& \twomodelNA{-1.91^{+0.41}_{-0.34}}\\

\midrule

$\Omega_{\rm m}$
& \twomodel{0.1857^{+0.0074}_{-0.043}}{0.3161\pm0.0072}
& \twomodel{0.355\pm0.026}{0.3121\pm0.0055}
& \twomodel{0.350^{+0.021}_{-0.023}}{0.3013\pm0.0035}
& \twomodel{0.266^{+0.026}_{-0.018}}{0.2977\pm0.0061}
& \twomodel{0.2830\pm0.0064}{0.3004^{+0.0057}_{-0.0064}}
& \twomodel{0.2792^{+0.0057}_{-0.0064}}{0.2903\pm0.0053}\\

$\Omega_{\rm m} h^2$
& \twomodel{0.1424\pm0.0011}{0.1431\pm0.0011}
& \twomodel{0.1432\pm0.0010}{0.14256\pm0.00086}
& \twomodel{0.14274^{+0.00084}_{-0.00076}}{0.14096\pm0.00060}
& \twomodel{0.1424\pm0.0010}{0.14040\pm0.00098}
& \twomodel{0.1421\pm0.0011}{0.14091^{+0.00090}_{-0.0010}}
& \twomodel{0.1419\pm0.0011}{0.13936\pm0.00088}\\

$w(z\rightarrow\infty)$
& \twomodelNA{-2.58^{+0.75}_{-1.3}}
& \twomodelNA{-1.99^{+0.50}_{-0.43}}
& \twomodelNA{-2.10^{+0.41}_{-0.35}}
& \twomodelNA{-2.48^{+0.38}_{-0.88}}
& \twomodelNA{-2.40^{+0.33}_{-0.28}}
& \twomodelNA{-2.57^{+0.34}_{-0.27}}\\

$q_0$
& \twomodel{-1.23^{+0.44}_{-0.55}}{-0.526\pm0.011}
& \twomodel{0.03\pm0.26}{-0.5318\pm0.0082}
& \twomodel{0.06\pm0.22}{-0.5481\pm0.0053}
& \twomodel{-0.39^{+0.47}_{-0.20}}{-0.5535\pm0.0091}
& \twomodel{-0.246\pm0.083}{-0.5495^{+0.0085}_{-0.0096}}
& \twomodel{-0.214\pm0.086}{-0.5645\pm0.0079}\\

$\sigma_8$
& \twomodel{0.986^{+0.076}_{-0.027}}{0.8124\pm0.0050}
& \twomodel{0.783\pm0.022}{0.8123\pm0.0053}
& \twomodel{0.784\pm0.017}{0.8109^{+0.0051}_{-0.0056}}
& \twomodel{0.865^{+0.022}_{-0.032}}{0.8098\pm0.0054}
& \twomodel{0.8421\pm0.0088}{0.8112\pm0.0054}
& \twomodel{0.8451\pm0.0094}{0.8100\pm0.0059}\\

$S_8$
& \twomodel{0.766^{+0.015}_{-0.032}}{0.834\pm0.012}
& \twomodel{0.850\pm0.013}{0.8285\pm0.0095}
& \twomodel{0.846\pm0.012}{0.8126\pm0.0075}
& \twomodel{0.812^{+0.015}_{-0.011}}{0.807\pm0.010}
& \twomodel{0.818\pm0.011}{0.8117^{+0.0096}_{-0.011}}
& \twomodel{0.815\pm0.011}{0.7968\pm0.0095}\\

$r_{\rm d}\,[\mathrm{Mpc}]$
& \twomodel{147.21\pm0.26}{147.07\pm0.26}
& \twomodel{147.04\pm0.24}{147.17\pm0.23}
& \twomodel{147.15\pm0.21}{147.48\pm0.19}
& \twomodel{147.21\pm0.25}{147.59\pm0.25}
& \twomodel{147.25\pm0.26}{147.45\pm0.24}
& \twomodel{147.29\pm0.26}{147.75\pm0.23}\\

$r_{\rm d} h$
& \twomodel{131^{+20}_{-5}}{98.99\pm0.90}
& \twomodel{93.6^{+3.3}_{-3.7}}{99.48\pm0.69}
& \twomodel{94.1\pm2.9}{100.88^{+0.44}_{-0.49}}
& \twomodel{108.0^{+3.2}_{-5.6}}{101.38\pm0.82}
& \twomodel{104.4\pm1.1}{101.01\pm0.80}
& \twomodel{105.0\pm1.0}{102.38\pm0.74}\\

\midrule

$\chi^2_{\min}$
& \twomodel{2826.36}{2827.66}
& \twomodel{2836.11}{2840.40}
& \twomodel{2834.34}{2843.69}
& \twomodel{2838.70}{2869.53}
& \twomodel{4291.31}{4317.18}
& \twomodel{4307.18}{4350.49}\\

$\Delta\chi^2_{\min}$
& \twomodel{-1.30}{0.00}
& \twomodel{-4.29}{0.00}
& \twomodel{-9.34}{0.00}
& \twomodel{-30.82}{0.00}
& \twomodel{-25.88}{0.00}
& \twomodel{-43.32}{0.00}\\

$\ln\mathcal{Z}$
& \twomodel{-1450.79\pm0.29}{-1451.24\pm0.29}
& \twomodel{-1460.60\pm0.29}{-1457.66\pm0.28}
& \twomodel{-1459.05\pm0.30}{-1459.53\pm0.29}
& \twomodel{-1460.95\pm0.29}{-1472.93\pm0.29}
& \twomodel{-2188.49\pm0.30}{-2196.48\pm0.29}
& \twomodel{-2197.12\pm0.30}{-2212.83\pm0.29}\\

$\Delta\ln\mathcal{Z}$
& \twomodel{0.45}{0.00}
& \twomodel{-2.94}{0.00}
& \twomodel{0.48}{0.00}
& \twomodel{11.98}{0.00}
& \twomodel{7.98}{0.00}
& \twomodel{15.71}{0.00}\\

\bottomrule

\end{tabular}%
}%
\end{table*}

\subsection{Constraints on Dark Energy Parameters and the Hubble Tension}
\label{subsecIIIA}

Before interpreting the CMB-only CPL constraints, we stress that they are broad and are dominated by the geometric degeneracy between $(H_0,\Omega_{\rm m},w_0,w_a)$. Extreme regions of the CMB-only posterior, including the apparent extension toward $q_0<-1$, should therefore be read together with the discussion in Sec.~III\,B, where we show that these features are degeneracy/extrapolation effects in the absence of low-redshift distance anchors.
\begin{figure}[t!]
  \centering
  \includegraphics[width=\linewidth]{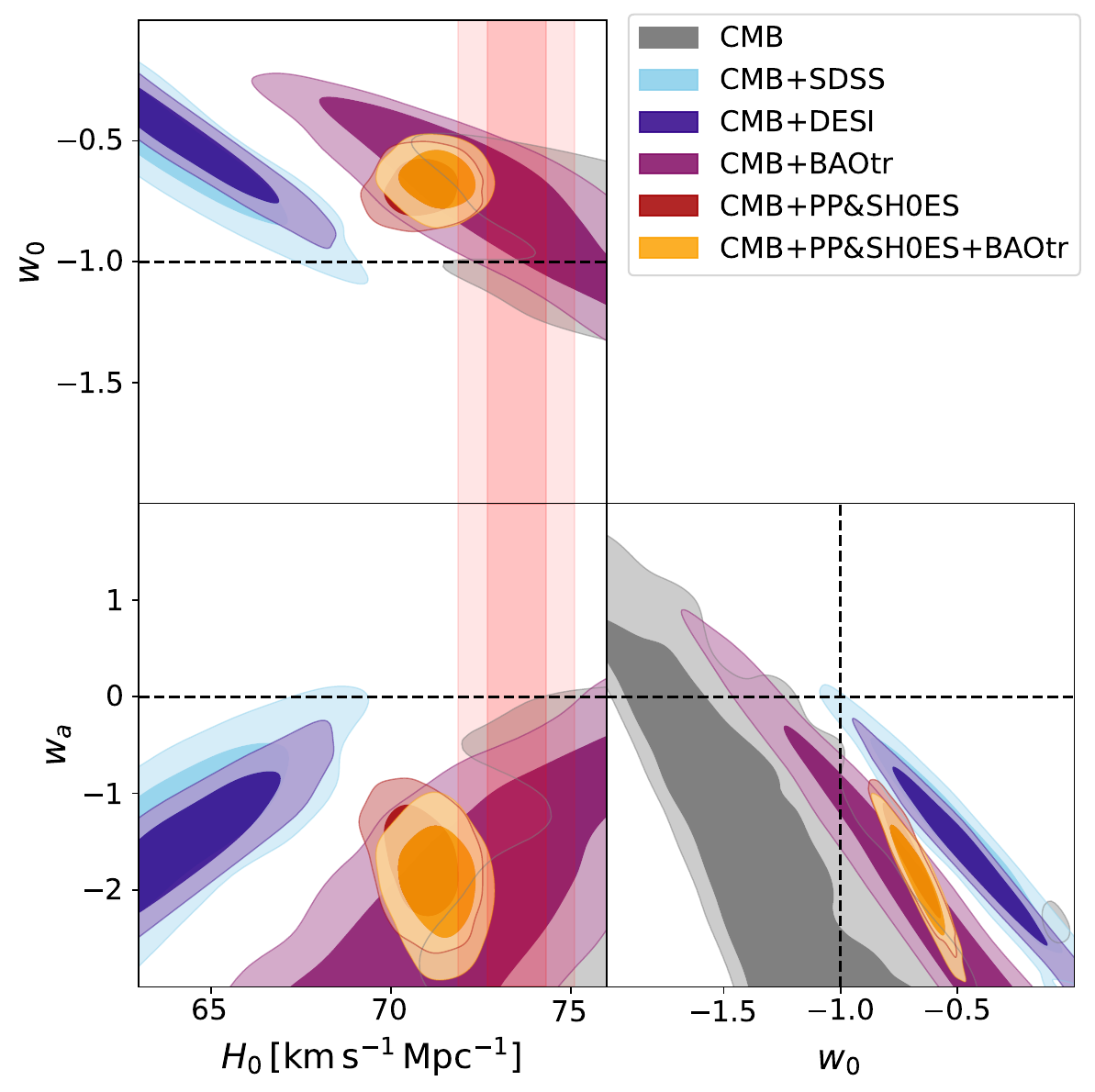}
  \caption{\label{fig:w0waH0_contours}
Two-dimensional marginalized posteriors in the $w_0w_a$CDM (CPL) model showing the correlations among $(w_0,w_a,H_0)$ for the dataset combinations listed in the legend. Panels show $(w_0,w_a)$ (left), $(w_0,H_0)$ (middle), and $(w_a,H_0)$ (right). The red vertical bands in the $H_0$ panels indicate the H0DN determination~\cite{H0DN:2025lyy} at $\pm1\sigma$ and $\pm2\sigma$ for reference. Contours enclose 68\% and 95\% credible regions.
}

\end{figure}

\cref{fig:w0waH0_contours} presents the posterior distributions of $(w_0,w_a)$ and their correlation with $H_0$ in the $w_0w_a$CDM model for the various dataset combinations. The CPL constraints are strongly dataset-dependent: combinations that include standard three-dimensional BAO distances (SDSS or DESI) prefer a low-$H_0$ solution and a distinct region in the $(w_0,w_a)$ plane, whereas combinations involving PP\&SH0ES and/or BAOtr shift the posterior toward higher $H_0$ and a different CPL locus. As a result, CMB+BAOtr and CMB+PP\&SH0ES are mutually consistent in the inferred CPL parameter space, while both are in $>2\sigma$ tension with the CMB+SDSS-inferred $(w_0,w_a)$ constraints. When the more precise three-dimensional BAO measurements from DESI are incorporated (CMB+DESI), the mismatch with CMB+PP\&SH0ES(+BAOtr) is further enhanced. Several of these combinations also disfavor the $\Lambda$CDM point $(w_0,w_a)=(-1,0)$ at the $\gtrsim2\sigma$ level within CPL, indicating a dataset-contingent preference for evolving dark energy rather than a single, dataset-independent CPL determination.

\begin{figure}[ht!]
  \centering
  \includegraphics[width=\linewidth]{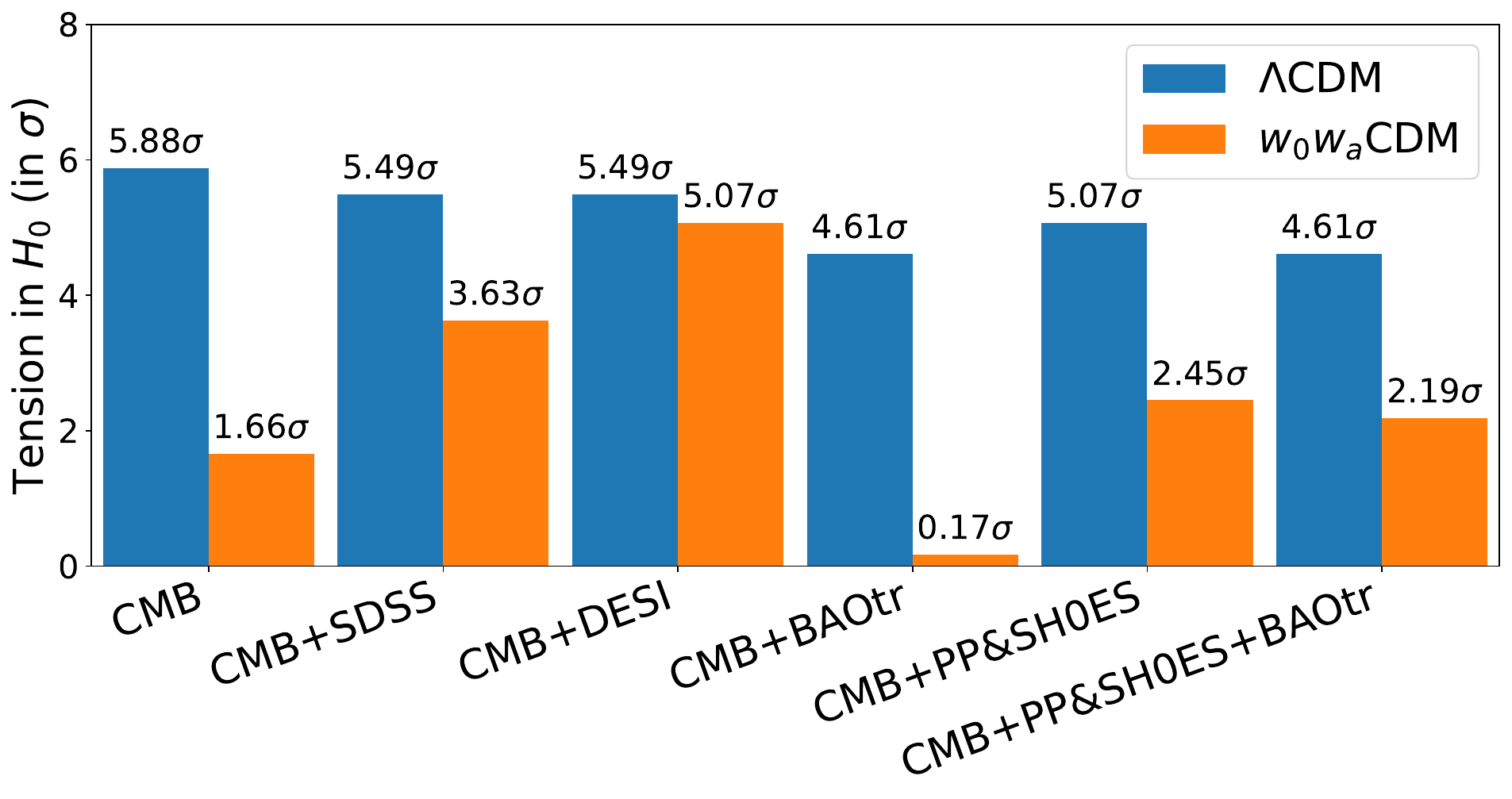}
  \caption{\label{fig:h0_tension}
Hubble-constant tension with respect to the H0DN determination, shown for each dataset combination in $\Lambda$CDM (left) and $w_0w_a$CDM (right). For each case, we form the difference distribution $\Delta H_0 = H_0^{\rm (case)} - H_0^{\rm (H0DN)}$ from posterior samples and report the equivalent Gaussian tension $T_\sigma$ (see~\cref{eqn:T}). Combinations involving PP\&SH0ES and/or BAOtr yield substantially reduced tension in $w_0w_a$CDM, while combinations with standard three-dimensional BAO (SDSS or DESI) remain in significant tension.}
\end{figure}

As seen from~\cref{fig:w0waH0_contours,fig:h0_tension}, allowing CPL freedom can substantially reduce the Hubble tension for specific late-time combinations, particularly when CMB is combined with PP\&SH0ES and/or BAOtr. For PP\&SH0ES-inclusive combinations, part of this reduction is expected because the likelihood already contains Cepheid-calibrated local SN~Ia distance-ladder information; the residual comparison with H0DN should therefore be interpreted as a consistency check with an external local-distance synthesis rather than as a fully independent local-data test. In contrast, when CMB is combined with standard three-dimensional BAO measurements (SDSS or DESI), the inferred $H_0$ is pulled toward lower values and the Hubble tension remains severe. This illustrates that the apparent CPL ``resolution'' of the Hubble tension is not generic, but depends sensitively on which late-time distance information is included.

\subsection{Late-time Expansion History and the Deceleration Parameter}\label{subsecIIIB}

\begin{figure}[b!]
  \centering
  \includegraphics[width=\linewidth]{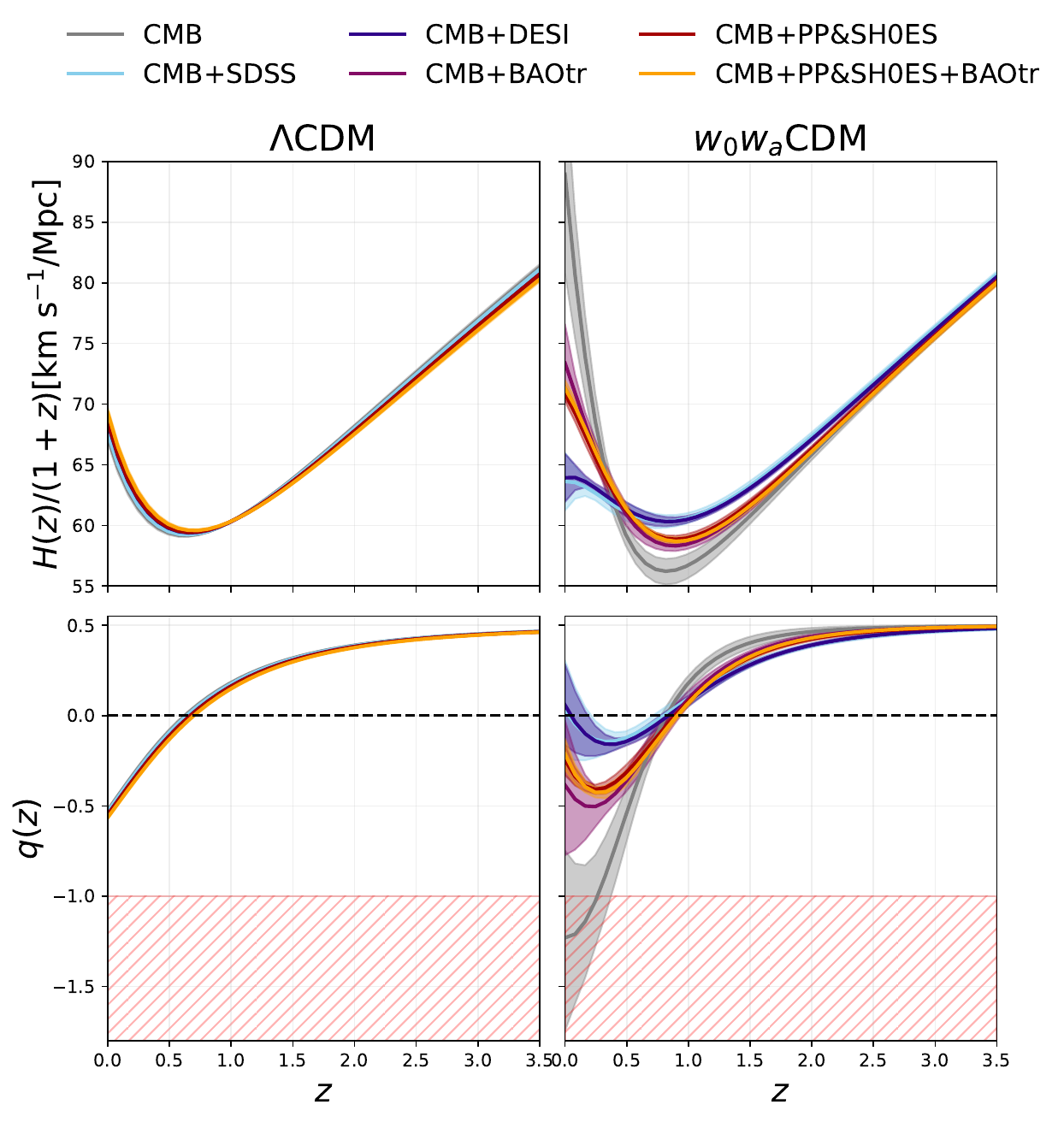}
  \caption{\label{fig:H_q_z}
Late-time expansion history reconstructed from each dataset combination in $\Lambda$CDM (left) and $w_0w_a$CDM (right). Top panels show the conformal Hubble rate $H(z)/(1+z)$; bottom panels show the deceleration parameter $q(z)$. Solid curves denote posterior means and shaded bands the $1\sigma$ credible regions. The horizontal dashed line marks the acceleration boundary $q=0$. The hatched region ($q<-1$) corresponds to super-acceleration ($\dot H>0$), i.e. $w_{\rm tot}<-1$ for the \emph{total} cosmic fluid; in GR this implies $\rho_{\rm tot}+p_{\rm tot}<0$ (NEC violation by the total cosmic fluid in GR). The strong spread among reconstructions in the $w_0w_a$CDM case highlights the pronounced dataset dependence of CPL late-time dynamics.
}
\end{figure}

\cref{fig:H_q_z} shows the late-time expansion histories---namely the Hubble rate and the deceleration parameter---for $\Lambda$CDM and $w_0w_a$CDM under the various dataset combinations. The deceleration parameter is defined as
\begin{equation}
\begin{aligned}
q(z) &\equiv -\frac{a\ddot{a}}{\dot{a}^2}
= -1-\frac{\dot H}{H^2}
= \frac{1}{2}\sum_{i}\Omega_i(z)\!\left[1+3w_i(z)\right]\\
&= \frac{1}{2}+\frac{3}{2}\sum_{i}\Omega_i(z)w_i(z), \quad
i\in \{\mathrm{m},\mathrm{r},\mathrm{DE},\ldots\},
\end{aligned}
\end{equation}
so that $q<0$ corresponds to accelerated expansion and $q>0$ to deceleration. Throughout, when we refer to the ``high-redshift'' behavior of $q(z)$ we mean higher redshift within the post-recombination range relevant to our late-time datasets (i.e.\ well below matter--radiation equality), for which the $\Lambda$CDM-like matter-dominated limit is $q\to 1/2$; the radiation-dominated limit $q\to 1$ at $z\gg z_{\rm eq}$ is standard and not probed here. Additionally, $q<-1$ implies super-acceleration ($\dot H>0$), i.e. $w_{\rm tot}< -1$ for the \emph{total} cosmic fluid. In GR this corresponds to $\rho_{\rm tot}+p_{\rm tot}<0$ (NEC violation by the \emph{total} cosmic fluid in GR). Importantly, as emphasized by Caldwell \& Linder~\cite{Caldwell:2025inn}, phantom-like behavior of an \emph{effective} dark-energy sector (e.g., $w_{\rm de}<-1$) does not by itself imply a fundamental pathology, since energy conditions apply to the total stress-\allowbreak energy rather than to an arbitrarily defined component.\footnote{
The implication $q<-1\iff \dot H>0 \Rightarrow \rho_{\rm tot}+p_{\rm tot}<0$ is a GR statement for an FLRW background, using the GR relation
$\dot H=-4\pi G(\rho_{\rm tot}+p_{\rm tot})$ (spatial curvature neglected).
In modified-gravity frameworks the background field equations are altered, and the same expansion history can be rewritten as GR with an \emph{effective} dark sector; in that case $q<-1$ does not automatically imply a fundamental instability and must be assessed at the level of perturbations.
A concrete example is type-II minimally modified gravity such as VCDM, which modifies the background while propagating only the two tensor modes (no extra propagating scalar) and can admit effective phantom-like background evolution without introducing a propagating ghost degree of freedom by construction~\cite{DeFelice:2020eju,DeFelice:2022uxv,Akarsu:2024qsi}.
Thus, if future data were to robustly require $q_0<-1$ beyond modeling/degeneracy effects, within GR it would indicate genuinely phantom-like behavior of the \emph{total} cosmic fluid, whereas in modified-gravity or nonstandard dark-sector frameworks it could instead point to new gravitational dynamics rather than an instability.} This distinction is particularly relevant here, because the phantom-like region arises only in the CMB-only CPL tail and disappears once low-redshift distance anchors are included, consistent with an inference-driven effect rather than a robust physical requirement.

\begin{figure}[t!]
  \centering
  \includegraphics[width=\linewidth]{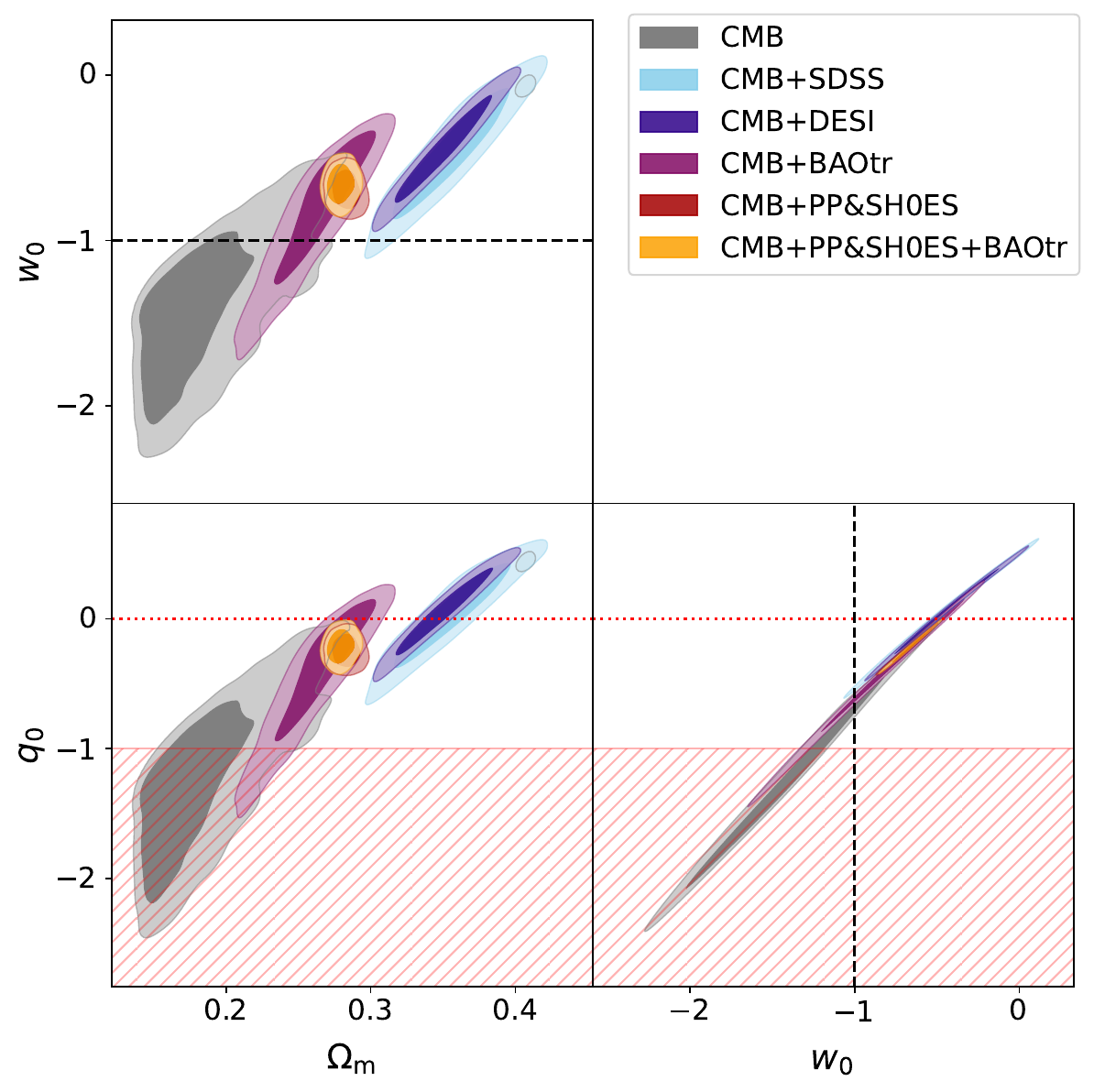}
  \caption{\label{fig:w0q0omegam}
Correlations among the present-day CPL parameter $w_0$, the deceleration parameter $q_0$, and the matter density $\Omega_{\rm m}$ in the $w_0w_a$CDM model for the dataset combinations considered. Different late-time distance probes select distinct regions of this parameter space. In particular, the CMB-only CPL posterior allows an extended phantom-like tail ($w_0<-1$) that maps to $q_0\lesssim -1$ when combined with low $\Omega_{\rm m}$, reflecting the broad CMB geometric degeneracy in CPL. By contrast, CMB+DESI favors higher $\Omega_{\rm m}$ and a nearly coasting present-day expansion ($q_0\simeq 0$). Contours enclose 68\% and 95\% credible regions.
}
\end{figure}

Within the $w_0w_a$CDM framework, different dataset combinations lead to markedly different late-time reconstructions for both $H(z)$ and $q(z)$. The present-day deceleration parameter may be approximated as
\begin{equation}
q_0 \equiv q(z=0) \simeq \frac{1}{2} + \frac{3}{2}(1-\Omega_{\rm m,0})w_0 ,
\end{equation}
so that shifts in $(w_0,\Omega_{\rm m,0})$ directly map into qualitatively different inferences about the present expansion state.

As seen from~\cref{fig:w0q0omegam} and~\cref{tab:parameters}, CMB-only constraints allow (and the posterior mean lies in) $w_0<-1$ together with a relatively low matter density ($\Omega_{\rm m,0}\lesssim0.2$), whereas including DESI DR2 BAO shifts the constraints toward $w_0\simeq-0.45$ and $\Omega_{\rm m,0}\simeq0.35$. Consequently, the CMB-only CPL posterior extends into $q_0\lesssim-1$, while CMB+DESI yields $q_0\simeq 0$, consistent with the trends in~\cref{fig:H_q_z}. 

This indicates that when CMB data are used alone, the inferred CPL parameters can drive the reconstructed deceleration parameter below $q_0=-1$, corresponding to super-acceleration ($\dot H>0$) and $w_{\rm tot}<-1$. However, in a flexible late-time model such as CPL, CMB-only constraints largely probe the distance to last scattering and admit a broad geometric degeneracy in $(H_0,\Omega_{\rm m},w_0,w_a)$; in this situation the inferred $q_0$ can become prior-volume/extrapolation dominated (as also reflected by the one-sided CMB-only constraint on $H_0$). We therefore interpret the $q_0<-1$ region in the CMB-only CPL fit as a degeneracy-driven CPL extrapolation artifact in the absence of low-redshift distance information, rather than as a robust inference of super-acceleration in the real Universe.

By contrast, the inclusion of DESI DR2 BAO data shifts the reconstructed expansion history toward $q_0 \simeq 0$, implying a marginally accelerating or nearly coasting Universe within the CPL framework. Rather than reflecting a definitive statement about the true cosmic expansion, this result highlights the strong sensitivity of the CPL parametrization to the choice of low-redshift datasets.

In comparison, the CMB+PP\&SH0ES+BAOtr combination yields parameter constraints that avoid these extreme behaviors, with $q_0$ remaining in the range $-1 < q_0 \lesssim 0$, corresponding to a moderately accelerating expansion. Taken together, these results support the conclusion that CPL reconstructions of late-time expansion can be unstable under mutually pulling low-redshift distance information, and that extreme inferences (such as $q_0\lesssim-1$) primarily arise in dataset combinations where late-time distances do not sufficiently anchor the background evolution.

Finally, the derived CPL high-redshift asymptote should be interpreted as an extrapolated descriptor of the CPL posterior rather than a direct high-redshift measurement or a direct constraint on early-dark-energy models. Specifically,
$w(z\to\infty)=w(a\to 0)=w_0+w_a$
is phantom-like ($< -1$) for all dataset combinations considered (see the $w(z\to\infty)$ row in~\cref{tab:parameters}), implying that the CPL dark-energy equation of state becomes more negative toward higher redshift in the region of parameter space selected by the data. Even when only an upper limit on $w_a$ is available, the posterior correlation between $(w_0,w_a)$ still drives $w_0+w_a$ to $<-1$. Despite this phantom-like dark-energy asymptote, the corresponding deceleration histories remain entirely non-phantom at high redshift. Indeed,~\cref{fig:H_q_z} shows that for $z\gtrsim 0.5$ (well within the post-recombination regime relevant to our datasets) the reconstructed deceleration parameter increases monotonically and approaches the $\Lambda$CDM-like matter-dominated limit $q\to 1/2$ in all cases, indicating that the dark-energy fraction is already negligible in this regime. Thus, phantom-like behavior of the \emph{dark-energy sector} at high redshift does not imply $w_{\rm tot}<-1$ for the \emph{total} cosmic fluid (or $q<-1$): the total expansion remains matter dominated at those redshifts, and super-acceleration requires the \emph{total} equation of state to cross below $-1$ rather than a component-level phantom asymptote~\cite{Caldwell:2025inn}. It is also worth noting that, once $w_0+w_a<-1$ is favored, the CPL fit implies a rapidly decreasing dark-energy density toward high redshift, since $\rho_{\rm de}\propto (1+z)^{3(1+w_0+w_a)}$ asymptotically. Within the standard-fluid interpretation of CPL this decay can only asymptote to $\rho_{\rm de}\to 0$ at early times; nevertheless, it motivates exploring sign-changing scenarios in which $\rho_{\rm de}(z)$ continues to decrease through zero and becomes negative at sufficiently high redshift, for example in late-time AdS-to-dS transition models such as $\Lambda_{\rm s}$CDM~\cite{Akarsu:2019hmw,Akarsu:2021fol,Akarsu:2022typ,Akarsu:2023mfb}, braneworld~\cite{Sahni:2002dx,Mishra:2025goj}, and teleparallel $f(T)$ gravity~\cite{Akarsu:2024nas}. These are physically distinct extensions motivated by the same qualitative data-driven trend, not continuations of CPL itself.

\subsection{Model Comparison and Bayesian Evidence}
\label{subsecIIIC}

\begin{figure}[t!]
  \centering
  \includegraphics[width=\linewidth]{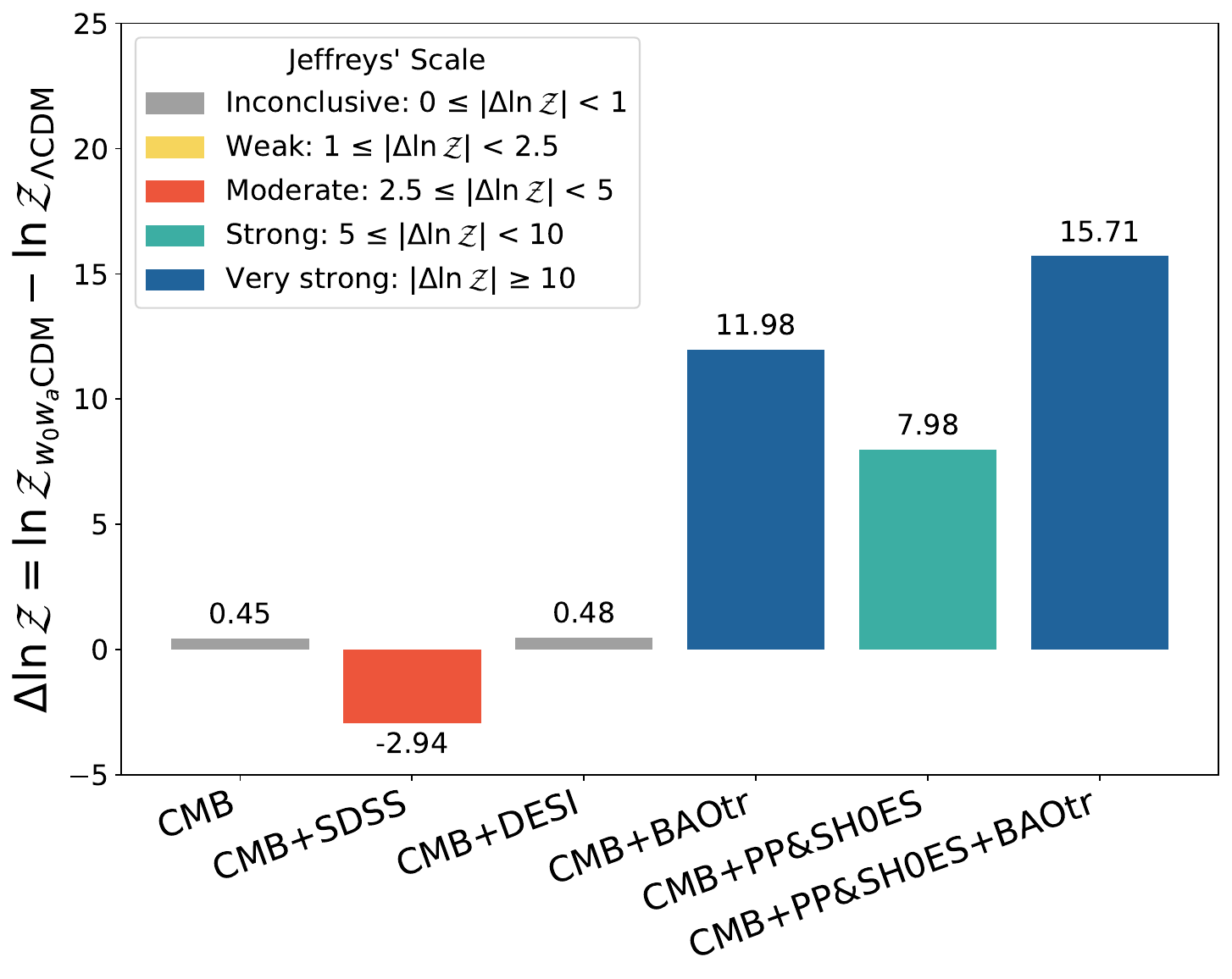}
  \caption{\label{fig:BE}
Bayesian model comparison between $w_0w_a$CDM (CPL) and $\Lambda$CDM for each dataset combination. Bars show the evidence difference
$\Delta\ln\mathcal{Z}\equiv \ln\mathcal{Z}(\mathrm{CPL})-\ln\mathcal{Z}(\Lambda\mathrm{CDM})$; $\Delta\ln\mathcal{Z}>0$ favors CPL. The revised Jeffreys' scale (Trotta) is overlaid to interpret the strength of evidence, as defined in the text. Strong-to-very-strong evidence in favor of CPL is obtained for combinations including PP\&SH0ES and/or BAOtr, while CMB-only and CMB+DESI remain inconclusive; CMB+SDSS yields moderate evidence in favor of $\Lambda$CDM.
}
\end{figure}

As a first goodness-of-fit diagnostic,~\cref{tab:parameters} shows that $w_0w_a$CDM (CPL) yields a lower best-fit $\chi^2_{\min}$ than $\Lambda$CDM for all dataset combinations, i.e. $\Delta\chi^2_{\min}\equiv \chi^2_{\min}(\mathrm{CPL})-\chi^2_{\min}(\Lambda\mathrm{CDM})<0$. This is not surprising: CPL introduces two additional parameters and therefore has the flexibility to improve the fit whenever there is residual structure in the late-time distance data that $\Lambda$CDM cannot absorb. As a complementary (approximate) model-selection check, one may consider the Akaike information criterion ${\rm AIC}\equiv \chi^2_{\min}+2k$, for which $\Delta{\rm AIC}=\Delta\chi^2_{\min}+4$ since CPL adds two parameters. Thus, AIC favors CPL whenever the fit improvement is substantial ($\Delta\chi^2_{\min}<-4$), which is readily satisfied for combinations including BAOtr and/or PP\&SH0ES, while it is not compelling for cases where the fit improvement is modest.

\begin{figure}[b!]
  \centering
  \includegraphics[width=\linewidth]{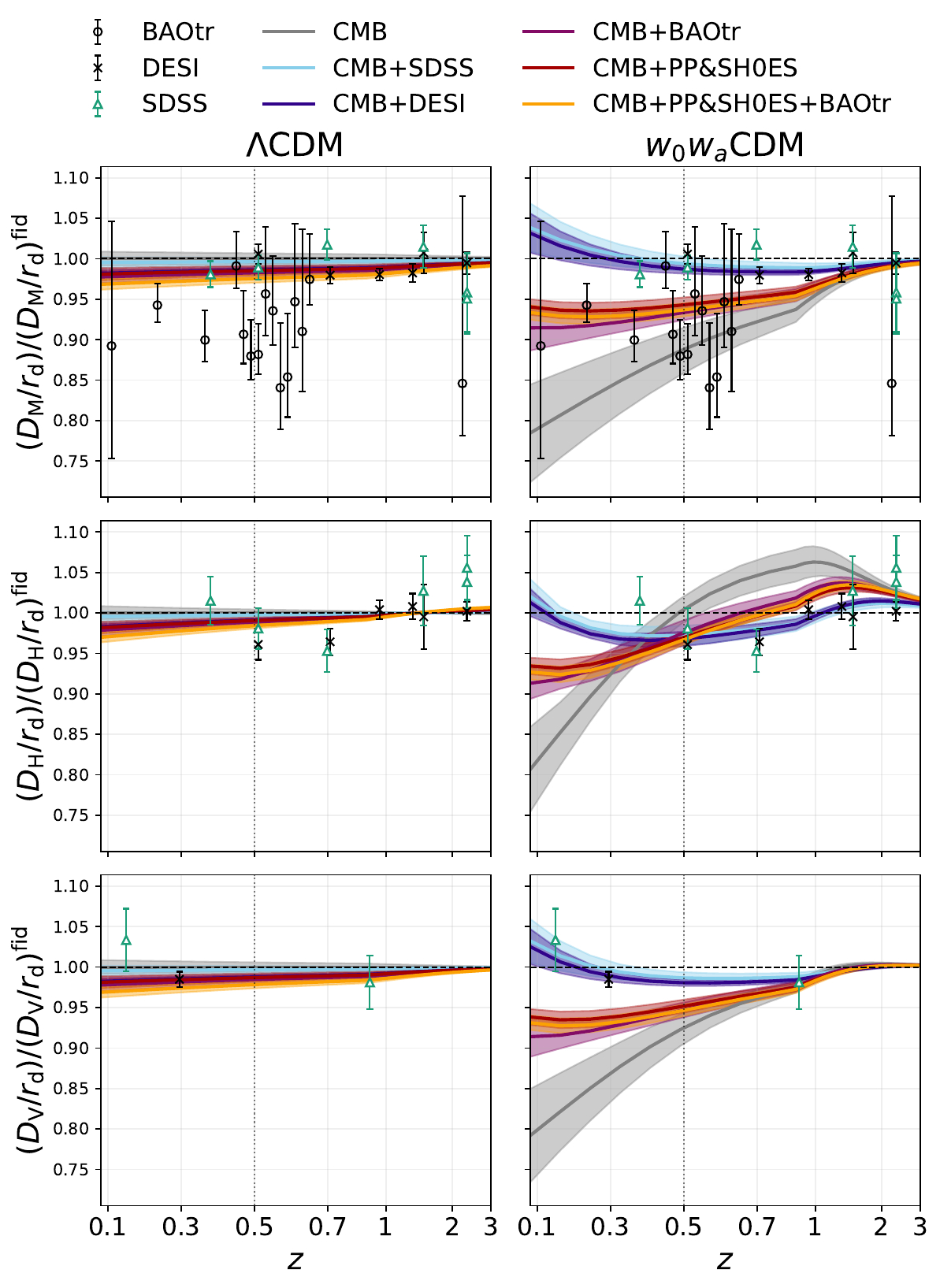}
  \caption{\label{fig:d_z}
Reconstructed BAO distance ratios $D_{\rm M}/r_{\rm d}$, $D_{\rm H}/r_{\rm d}$, and $D_{\rm V}/r_{\rm d}$ for $\Lambda$CDM (left) and $w_0w_a$CDM (right). DESI DR2, SDSS-IV/eBOSS, and BAOtr measurements are shown as crosses, open triangles, and open circles; curves/bands give posterior means and $1\sigma$ regions, normalized to the CMB-only $\Lambda$CDM best-fit ratios for visual comparison. This normalization is a plotting convention only and does not assign preferential physical status to the CMB-only $\Lambda$CDM best fit. The redshift axis is mixed linear-log (linear below $z\simeq0.8$), so the apparent bend near the transition is a coordinate artifact. DESI/SDSS error bars are diagonal or representative $1\sigma$ values from Appendix~\ref{app:bao_data}; the likelihoods use the full covariances or public grids where applicable. The CPL panels show the low-redshift BAO differences that drive dataset-dependent late-time reconstructions.
}
\end{figure}

\begin{figure}[b!]
  \centering
  \includegraphics[width=\linewidth]{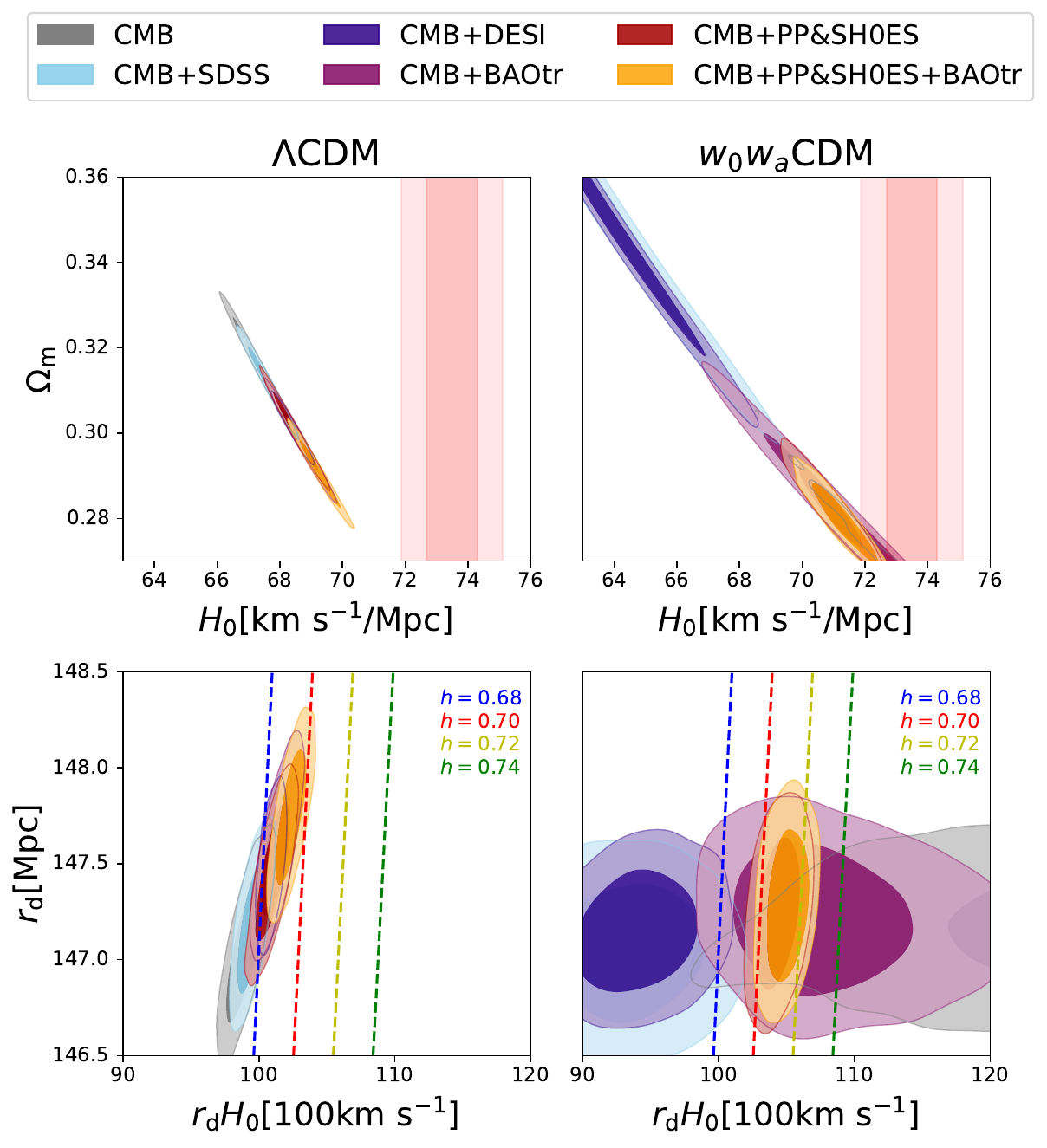}
\caption{\label{fig:rdH0Omegam}
Constraints on $H_0$, $\Omega_{\rm m}$, and sound-horizon-related combinations in $\Lambda$CDM (left) and $w_0w_a$CDM (right). Top panels show the $(H_0,\Omega_{\rm m})$ constraints, while bottom panels show $(r_{\rm d}H_0,r_{\rm d})$. In $\Lambda$CDM, $r_{\rm d}H_0$ is tightly constrained and nearly consistent across dataset combinations, yielding a narrow range of $H_0$ and maintaining substantial tension with local distance-ladder determinations. In CPL, where pre-recombination physics is fixed, $r_{\rm d}$ remains stable while $r_{\rm d}H_0$ (and hence $H_0$) shifts significantly with the choice of low-redshift distance data; the inferred changes in $H_0$ are therefore absorbed by late-time expansion-history freedom rather than by a change in the sound horizon. Dashed lines in the $r_{\rm d}$--$r_{\rm d}H_0$ panels denote constant $H_0$ with $h\equiv H_0/100\rm \,km\,s^{-1}\,Mpc^{-1}$.}
\end{figure}

However, the most stringent assessment comes from Bayesian evidence, which automatically accounts for model complexity through the prior volume. We compute $\ln\mathcal{Z}$ with \texttt{PolyChord} and define the Bayes factor $\ln\mathcal{B}_{\rm CPL,\Lambda{\rm CDM}}\equiv \Delta\ln\mathcal{Z}
= \ln\mathcal{Z}(\mathrm{CPL})-\ln\mathcal{Z}(\Lambda\mathrm{CDM})$, so that $\Delta\ln\mathcal{Z}>0$ favors CPL. We interpret $|\Delta\ln\mathcal{Z}|$ using the revised Jeffreys' scale of Trotta~\cite{Kass:1995loi,Trotta:2008qt} (see~\cref{fig:BE}). \cref{fig:BE} shows that the evidence is strongly dataset-dependent. For CMB-only and CMB+DESI, $\Delta\ln\mathcal{Z}\approx 0.5$, i.e. inconclusive evidence once the extended CPL parameter volume is accounted for. In contrast, when BAOtr and/or PP\&SH0ES are included, the evidence becomes strong to very strong in favor of CPL (e.g., $\Delta\ln\mathcal{Z}\gtrsim 8$ for PP\&SH0ES and $\Delta\ln\mathcal{Z}\gtrsim 12$ for BAOtr, reaching $\Delta\ln\mathcal{Z}\approx 15.7$ for CMB+PP\&SH0ES+BAOtr). The only case that moderately prefers $\Lambda$CDM is CMB+SDSS, for which $\Delta\ln\mathcal{Z}\approx -3$.

Bayesian evidence is necessarily prior-dependent. Our CPL runs use the broad flat priors listed in~\cref{tab:prior}, $w_0\in[-3,1]$ and $w_a\in[-3,2]$, together with $w_0+w_a<0$. These choices follow the DESI DR2 setup and are intentionally conservative, but they also impose an Occam penalty on CPL whenever the data do not localize the extra parameter volume. Consequently, the numerical values of $\Delta\ln\mathcal{Z}$, especially in the inconclusive CMB-only and CMB+DESI cases, should not be interpreted independently of the adopted prior ranges. The qualitative conclusion relevant here is the relative dataset dependence: the same priors give inconclusive or $\Lambda$CDM-favoring evidence for CMB-only/DESI/SDSS combinations, but strong evidence for CPL when PP\&SH0ES and/or BAOtr are included.

Overall, the contrast between ``always-improving'' $\chi^2_{\min}$ and the highly non-universal $\Delta\ln\mathcal{Z}$ highlights the central point: the apparent statistical support for CPL is not generic, but depends sensitively on which low-redshift distance information is included. In the next subsection (\cref{subsecIIID}) we investigate the origin of this behavior by examining the consistency among BAO datasets and how low-redshift distance ratios propagate into divergent late-time reconstructions in flexible dark-energy models.

\subsection{\texorpdfstring{Low-redshift BAO consistency and dataset dependence in the $w_0w_a$CDM model}{Low-redshift BAO consistency and dataset dependence in the w0waCDM model}}\label{subsecIIID}

The strong dataset dependence of the CPL evidence and late-time reconstructions suggests that different low-redshift distance probes are not simultaneously accommodated by a single CPL background evolution. Here we diagnose the origin of this behavior by examining how different BAO datasets constrain distance ratios and how these constraints propagate into $r_{\rm d}H_0$ and the inferred shape of $E(z)\equiv H(z)/H_0$.

The BAO observables can be written as
\begin{align}
  \frac{D_{\rm H}(z)}{r_{\rm d}} &= \frac{c}{r_{\rm d} H_0\,E(z)} , \\
  \frac{D_{\rm M}(z)}{r_{\rm d}} &= \frac{c}{r_{\rm d} H_0}\int_0^z\frac{dz'}{E(z')},
\end{align}
and their combination
\begin{equation}
D_{\rm V}(z)=\left[zD_{\rm H}(z)D_{\rm M}^2(z)\right]^{1/3}.
\end{equation}
These relations make explicit that BAO measurements constrain the combination $r_{\rm d}H_0$ together with the redshift dependence of $E(z)$; therefore, inconsistencies among BAO distance ratios at low redshift propagate directly into divergent late-time reconstructions in flexible models such as CPL.

\subsubsection{Data-level BAOtr--3D BAO comparison}

We first quantify the BAOtr--3D BAO mismatch directly at the data level, before invoking any cosmological fit. Since BAOtr measures the angular scale $\theta_{\rm BAO}=r_{\rm d}/D_{\rm M}$, we use $\theta_{\rm BAO}$ as the primary comparison variable. A published 3D transverse BAO measurement of $D_{\rm M}/r_{\rm d}$ implies the equivalent angular scale
\begin{equation}
\theta_{\rm 3D}(z)=\frac{180^\circ/\pi}{(D_{\rm M}/r_{\rm d})_{\rm 3D}},
\quad
\sigma_{\theta,{\rm 3D}}
= \theta_{\rm 3D}
\frac{\sigma_{D_{\rm M}/r_{\rm d}}}{(D_{\rm M}/r_{\rm d})_{\rm 3D}},
\end{equation}
which requires no numerical fiducial cosmology. We then define the single-bin tension
\begin{equation}
T_\theta(z_i)=
\frac{\theta^{\rm BAOtr}_i-\theta^{\rm 3D}_i}
{\sqrt{\sigma^2_{\theta,{\rm BAOtr},i}
+\sigma^2_{\theta,{\rm 3D},i}}}.
\end{equation}
For multi-bin comparisons we form the difference vector $\Delta_i=\theta^{\rm BAOtr}_i-\theta^{\rm 3D}_i$ and use
\begin{equation}
\chi^2_{\rm diff}=\Delta^T C_{\rm diff}^{-1}\Delta,
\qquad
C_{\rm diff}=C_{\rm BAOtr}+C_{\rm 3D},
\end{equation}
with the PTE computed for the corresponding number of bins and then converted to an equivalent two-sided Gaussian significance.
Since public cross-covariances between the BAOtr measurements and the corresponding 3D BAO determinations are not available, we neglect such cross-correlations in this diagnostic comparison.
The quoted $T_\theta$ and $\chi^2_{\rm diff}$ values should therefore be interpreted as data-level pulls rather than as a full joint-likelihood inconsistency.

The cleanest comparison occurs at $z=0.510$, where BAOtr and both DESI and SDSS constrain the same transverse BAO scale. BAOtr gives $\theta_{\rm BAO}=4.81\pm0.17^\circ$, while DESI LRG1 gives $D_{\rm M}/r_{\rm d}=13.588\pm0.168$, i.e. $\theta_{\rm 3D}=4.217\pm0.052^\circ$. This corresponds to $T_\theta=3.34\sigma$. Equivalently, writing the BAOtr point as $D_{\rm M}/r_{\rm d}=11.912\pm0.421$ gives a reciprocal-distance tension of $3.70\sigma$, in agreement with the independent analysis of Ref.~\cite{Pantos:2026baotr}. The corresponding SDSS DR12 transverse point at the same redshift gives $D_{\rm M}/r_{\rm d}=13.366\pm0.204$, or $\theta_{\rm 3D}=4.287\pm0.065^\circ$, yielding $T_\theta=2.87\sigma$ (or $3.11\sigma$ in the reciprocal-distance representation).

At nearby but non-identical redshifts we use a local, data-level interpolation of the BAOtr angular measurements. For a 3D transverse point at $z$ bracketed by BAOtr points $(z_1,\theta_1)$ and $(z_2,\theta_2)$, we set $\theta(z)=(1-t)\theta_1+t\theta_2$, with $t=(z-z_1)/(z_2-z_1)$, and propagate the BAOtr covariance as $C^{\rm BAOtr}_{ij}=\sum_k w_{ik}w_{jk}\sigma_{\theta,k}^2$. This assumes independent BAOtr redshift bins, since no full BAOtr covariance matrix is available. We do not extrapolate and require the bracketing interval to satisfy $\Delta z\leq0.15$. With this prescription, the SDSS transverse points at $z=0.380$ and $0.510$ give $\chi^2_{\rm diff}=12.35$ for two bins (PTE $=0.0021$, equivalent to $3.08\sigma$). The DESI transverse comparison reduces to the direct $z=0.510$ case under the same rule.

\begin{figure*}[t!]
  \centering
  \includegraphics[width=0.82\textwidth]{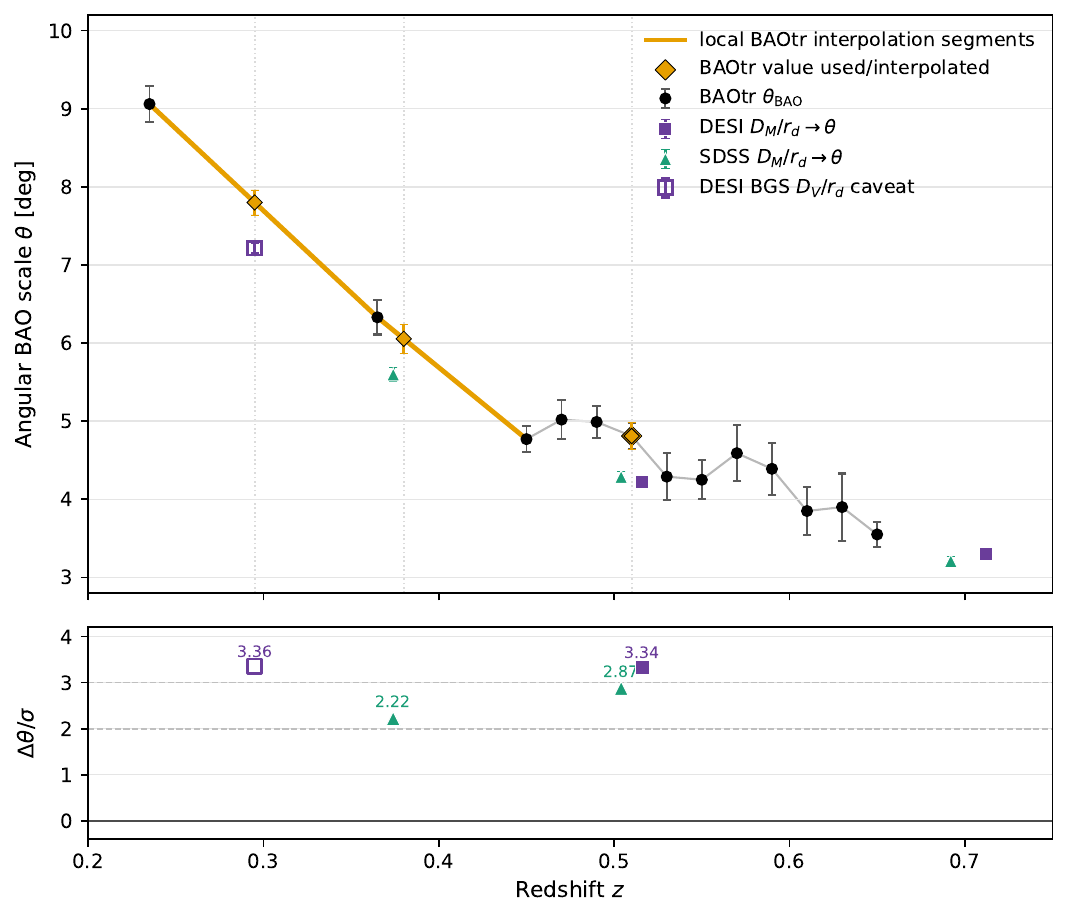}
  \caption{\label{fig:baotr_interpolation_visualization}
Visualization of the low-redshift BAOtr interpolation diagnostic used in~\cref{tab:bao_tension_data}. The upper panel shows the BAOtr angular measurements together with the local linear BAOtr interpolation segments used to compare with nearby 3D BAO transverse redshifts; no extrapolation is used, and only bracketing BAOtr intervals with $\Delta z\leq0.15$ are retained. DESI and SDSS transverse measurements are converted from $D_{\rm M}/r_{\rm d}$ to the equivalent angular scale $\theta=(180^\circ/\pi)/(D_{\rm M}/r_{\rm d})$. DESI and SDSS points at the same redshift are slightly offset horizontally for visual clarity. The DESI BGS point is shown separately as a caveated $D_{\rm V}/r_{\rm d}$ check and is not included in the primary transverse statistic. The lower panel shows the corresponding signed single-bin tension (pull), $\Delta\theta/\sigma$, in units of the combined $1\sigma$ uncertainty; global consistency is assessed separately with $\chi^2_{\rm diff}$.}
\end{figure*}

As a broader check, we also interpolate the 3D transverse anchors to the BAOtr redshifts covered by DESI or SDSS. Here the interpolated quantity is the published $X(z)\equiv D_{\rm M}(z)/r_{\rm d}$, with covariance $C_X^{\rm int}=W C_X W^T$, followed by the transformation $\theta=(180^\circ/\pi)/X$ using the linearized Jacobian. The resulting bin count therefore refers to BAOtr redshift points inside the 3D-anchor range, not to independent same-redshift 3D BAO measurements. Since no additional interpolation-model uncertainty is assigned, these broader results should be read as approximate correlated residual-vector checks. They give $\chi^2_{\rm diff}=19.68$ for 9 DESI--BAOtr bins (PTE $=0.020$, equivalent to $2.33\sigma$) and $\chi^2_{\rm diff}=22.92$ for 11 SDSS--BAOtr Gaussian transverse bins (PTE $=0.018$, equivalent to $2.36\sigma$). The DESI BGS point at $z=0.295$ measures $D_{\rm V}/r_{\rm d}$ rather than $D_{\rm M}/r_{\rm d}$ and is therefore kept outside the primary transverse statistic. As a caveated isotropic check, it gives a $3.36\sigma$ offset in the same qualitative low-redshift direction. The direct $z=0.510$ BAOtr--DESI transverse tension is independent of BGS. Thus, although the BGS point lies in the same low-redshift direction, our primary BAOtr--3D BAO conclusion does not rely on it; the direct DESI LRG1 comparison at $z=0.510$ and the SDSS cross-check already show the low-redshift transverse mismatch. These data-level offsets show that the BAOtr--DESI/SDSS difference is already present in the directly compared low-redshift BAO observables. They therefore provide a model-independent origin for the dataset dependence of the CPL reconstructions discussed below, rather than being merely a consequence of the late-time parametrization. The interpolation construction and the corresponding single-bin offsets are illustrated in~\cref{fig:baotr_interpolation_visualization} and summarized in~\cref{tab:bao_tension_data}.

\begin{table*}[t!]
\centering
\caption{\label{tab:bao_tension_data}
Data-level comparison between BAOtr angular measurements and the angular scale implied by 3D BAO transverse distances. The direct rows use matching redshifts; the interpolated row uses local linear interpolation of BAOtr in $z$ with no extrapolation and $\Delta z\leq0.15$. The DESI BGS row is shown only as a caveated isotropic-distance check because BGS measures $D_{\rm V}/r_{\rm d}$, not $D_{\rm M}/r_{\rm d}$. Possible cross-covariances between BAOtr and the 3D BAO measurements are not included because no public cross-covariance matrix is available; the quoted $T_\theta$ values should therefore be read as diagnostic data-level pulls.}

\setlength{\tabcolsep}{5pt}
\renewcommand{\arraystretch}{1.08}
\footnotesize
\begin{tabular}{llcccc}
\toprule
3D BAO comparison & Type & $z$ & BAOtr angle [deg] & 3D-implied angle [deg] & $T_\theta$ \\
\midrule
DESI LRG1 & direct $D_{\rm M}$ & 0.510 & $4.810\pm0.170$ & $4.217\pm0.052$ & 3.34 \\
SDSS DR12 galaxies & direct $D_{\rm M}$ & 0.510 & $4.810\pm0.170$ & $4.287\pm0.065$ & 2.87 \\
SDSS DR12 galaxies & local BAOtr interp. $D_{\rm M}$ & 0.380 & $6.055\pm0.184$ & $5.599\pm0.093$ & 2.22 \\
DESI BGS & caveated $D_{\rm V}$ check & 0.295 & $7.800\pm0.160$ & $7.215\pm0.069$ & 3.36 \\
\bottomrule
\end{tabular}
\end{table*}

\subsubsection{Fiducial-cosmology dependence of 3D BAO distances}

The above comparison also clarifies the role of the fiducial cosmology used in standard 3D BAO analyses. In a 3D BAO measurement, a fiducial model is required to convert observed angles and redshifts into comoving coordinates and to define the template against which the BAO feature is fitted. However, the published BAO distance ratios are not simply the fiducial distances. They are reported through Alcock--Paczynski dilation parameters, for example
\begin{equation}
\left(\frac{D_{\rm M}}{r_{\rm d}}\right)_{\rm pub}
=\alpha_\perp
\left(\frac{D_{\rm M}^{\rm fid}}{r_{\rm d}^{\rm fid}}\right),
\qquad
\alpha_\perp\simeq
\frac{D_{\rm M}^{\rm true}/r_{\rm d}^{\rm true}}
     {D_{\rm M}^{\rm fid}/r_{\rm d}^{\rm fid}},
\end{equation}
and analogously for the radial and isotropic combinations. Thus the leading dependence on the fiducial distance scale cancels in the reported quantity, up to residual effects associated with template shape, reconstruction, broadband modeling, and finite mock calibration~\cite{DESI:2025zgx}. Consequently, multiplying the published DESI or SDSS points by a ratio such as $D_{\rm M}^{\rm CPL}/D_{\rm M}^{\Lambda{\rm CDM}}$ would represent a full remapping of the coordinate system rather than a correction for the residual fiducial bias of the published BAO observable.

This distinction is important when interpreting the comparison between BAOtr and 3D BAO. The direct $z=0.510$ tension quoted above is formed from the published transverse distance ratio and the directly observed BAOtr angle, and therefore does not require choosing a numerical fiducial cosmology at the comparison stage. Residual fiducial effects in modern 3D BAO analyses are expected to be sub-percent, whereas the BAOtr--DESI offset at $z=0.510$ corresponds to several percent in $D_{\rm M}/r_{\rm d}$. As a numerical scale check, we compute a naive full-background remapping diagnostic using the mean CPL parameters in~\cref{tab:parameters}. Comparing to the CMB-only $\Lambda$CDM reference at $z=0.510$ gives $D_{\rm M}^{\rm CPL}/D_{\rm M}^{\Lambda{\rm CDM}}-1\simeq -5.5\%$ for CMB+PP\&SH0ES and $\simeq -6.1\%$ for CMB+PP\&SH0ES+BAOtr. These percentages are the bracketed factor that would multiply a published distance ratio in a naive rescaling estimate. They are comparable to the observed BAOtr--3D BAO difference, but they are not the actual fiducial-induced bias of the published BAO observable; they correspond to remapping the full background distance--redshift relation to a different cosmology. Determining the true residual fiducial bias would require a pipeline-level reanalysis in which the clustering measurement, reconstruction, template fitting, mock calibration, and likelihood construction are repeated with a CPL fiducial cosmology. The recent analysis of Ref.~\cite{Pantos:2026baotr} addresses this pipeline-level issue and finds residual fiducial effects at the $\lesssim0.1$--$0.3\%$ level. This is far smaller than the several-percent, $3.34\sigma$ BAOtr--DESI offset at $z=0.510$. We therefore regard the fiducial-cosmology issue as an important systematic check, but not as a sufficient explanation of the low-redshift BAOtr--DESI/SDSS mismatch.

\begin{figure}[t!]
  \centering
  \includegraphics[width=1.0\linewidth]{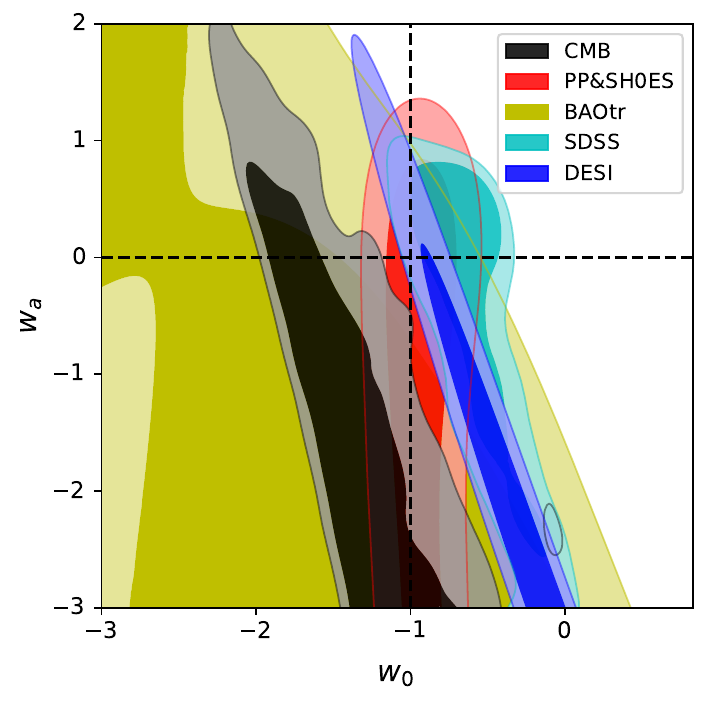}
\caption{\label{fig:w0wacdm_w0wa}
Constraints on the CPL parameters $(w_0,w_a)$ from individual probes (CMB, PP\&SH0ES, BAOtr, SDSS, and DESI DR2 BAO) within the $w_0w_a$CDM model. Contours enclose 68\% and 95\% credible regions; the $\Lambda$CDM point $(w_0,w_a)=(-1,0)$ is shown for reference. The preferred regions do not fully overlap, and in particular CMB and DESI DR2 BAO exhibit a $\gtrsim2\sigma$ mismatch in the inferred CPL parameter space. SDSS and DESI are broadly consistent with each other and prefer a narrow degeneracy band offset from the CMB-preferred region.}
\end{figure}

\cref{fig:d_z} compares DESI DR2 BAO, SDSS-IV/eBOSS BAO, and BAOtr measurements with the reconstructed BAO distance ratios in $\Lambda$CDM and $w_0w_a$CDM, shown as ratios normalized to the CMB-only $\Lambda$CDM best-fit values (``fid''). In $w_0w_a$CDM, the reconstructions diverge most strongly at low redshift ($z\lesssim 0.5$): BAOtr prefers slightly lower $D_{\rm M}/r_{\rm d}$, while DESI DR2 BAO shows a comparatively high $D_{\rm V}/r_{\rm d}$ point at $z=0.295$. By $z\sim 1$ the CPL reconstructions largely reconverge, consistent with the fact that BAO/SN constraints are most discriminating at low redshift for this comparison, highlighting that the dataset dependence is driven mainly by low-$z$ BAO distance information. In $\Lambda$CDM, the more restricted background evolution yields much smaller variation across dataset combinations, with only mild deviations emerging at $z\lesssim 1$.

\begin{figure*}[t!]
  \centering
  \includegraphics[width=0.49\linewidth]{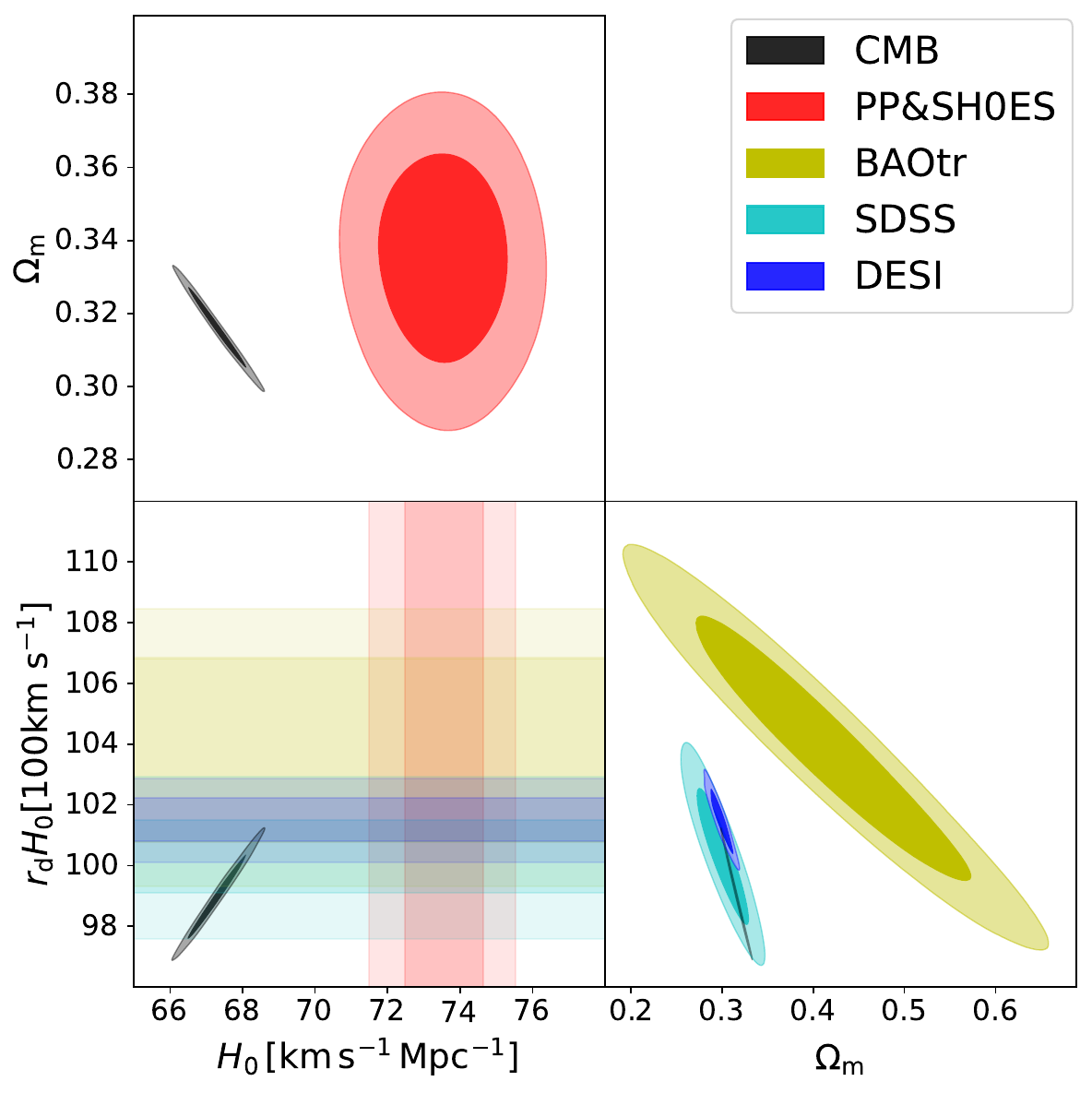}
  \includegraphics[width=0.49\linewidth]{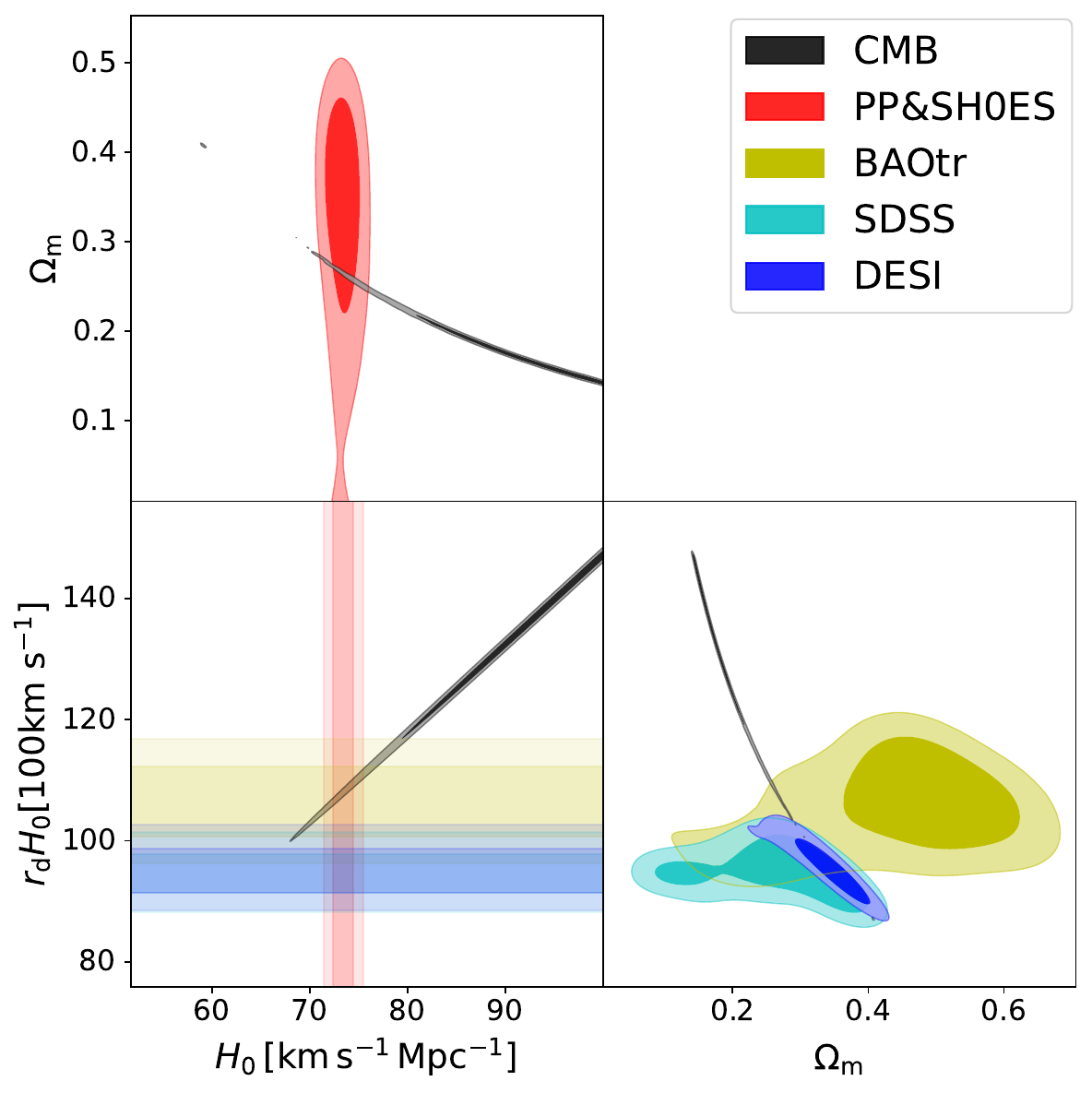}

\caption{\label{fig:individual_contour}
Consistency of individual probes under $\Lambda$CDM (left) and $w_0w_a$CDM (right) in representative parameter planes. Each panel overlays constraints from CMB, PP\&SH0ES, BAOtr, SDSS, and DESI DR2 BAO. In the $H_0$--$r_{\rm d}H_0$ plane, PP\&SH0ES appears as a vertical band (constraining $H_0$), while BAO measurements appear as horizontal bands (constraining $r_{\rm d}H_0$); CMB provides the joint contour constraints. In $\Lambda$CDM, CMB and BAO constraints are mutually consistent but remain in tension with PP\&SH0ES, reflecting the Hubble tension. In CPL, the additional late-time freedom allows CMB, BAOtr, and PP\&SH0ES to overlap, whereas SDSS/DESI prefer substantially lower $r_{\rm d}H_0$ (and lower $H_0$), removing a common intersection and highlighting an internal inconsistency among late-time distance probes within the CPL framework.}
\end{figure*}

\cref{fig:rdH0Omegam} summarizes how different low-redshift datasets map into $(H_0,\Omega_{\rm m})$ and $(r_{\rm d}H_0,r_{\rm d})$. In $\Lambda$CDM, standard pre-recombination physics fixes $r_{\rm d}$ and the CMB constrains $\Omega_{\rm m}h^2$, leaving a narrow allowed range of $H_0$. In CPL, the main dataset dependence appears instead in $r_{\rm d}H_0$ and hence in $H_0$, with the largest separation between CMB+DESI and combinations involving PP\&SH0ES and/or BAOtr. Quantitatively, the CPL mean values of $r_{\rm d}$ in~\cref{tab:parameters} span only $147.04$--$147.29\,{\rm Mpc}$, a fractional range of $0.17\%$, comparable to the individual $1\sigma$ errors. Across both CPL and $\Lambda$CDM the range remains below $0.5\%$. Thus the spread in $r_{\rm d}H_0$ mainly reflects changes in $H_0$ and in the reconstructed late-time expansion history $E(z)$, not a measured shift of the sound horizon. This statement is internal to the fixed-pre-recombination-physics CPL setup and should not be read as a direct constraint on EDE or on other models that alter the early sector.

To assess the consistency among individual probes within the CPL framework, we show in~\cref{fig:w0wacdm_w0wa} the two-dimensional constraints on $(w_0,w_a)$ obtained from the individual datasets (CMB, PP\&SH0ES, BAOtr, SDSS, and DESI DR2 BAO). These contours are not mutually overlapping: while BAOtr provides relatively broad constraints in the CPL plane, the standard three-dimensional BAO datasets (SDSS and DESI) prefer a narrow degeneracy band that is offset from the CMB-preferred region, yielding a $\gtrsim2\sigma$ mismatch between CMB and DESI DR2 BAO in the inferred CPL parameter space.

A complementary view is provided by~\cref{fig:individual_contour}, which shows the corresponding constraints in the $(H_0,\Omega_{\rm m})$ and $(r_{\rm d}H_0,H_0)$ planes under $\Lambda$CDM and CPL. Since PP\&SH0ES does not constrain $r_{\rm d}H_0$ and BAO data alone do not constrain $H_0$, we represent their constraints in the $r_{\rm d}H_0$--$H_0$ plane as vertical (PP\&SH0ES) and horizontal (BAO) bands, respectively. In the $\Lambda$CDM framework, the CMB and BAO datasets (BAOtr/SDSS/DESI) are mutually consistent but exhibit a clear tension with PP\&SH0ES, reflecting the Hubble tension. In the CPL framework, the additional late-time freedom allows CMB, PP\&SH0ES, and BAOtr to achieve overlapping regions in the $r_{\rm d}H_0$--$H_0$ plane, whereas SDSS and DESI prefer substantially lower $r_{\rm d}H_0$ (and correspondingly lower $H_0$) and occupy a distinct region that removes any common intersection among the three. Notably, SDSS and DESI are broadly consistent with each other in CPL, so the dominant mismatch is between BAOtr and the standard three-dimensional BAO determinations at low redshift. Taken together, these results indicate that, assuming the data are free from significant systematics, the two-parameter CPL form may be too restrictive to accommodate all late-time distance information simultaneously.

\section{Conclusion}
\label{sec:conclusion}

In this study, we have conducted a comprehensive analysis of the late-time expansion history within the standard $\Lambda$CDM paradigm and its canonical dynamical extension, the Chevallier--Polarski--Linder (CPL) parametrization ($w_0w_a$CDM)~\cite{Chevallier:2000qy,Linder:2002et}. By combining early-Universe information from CMB anisotropies and lensing with a suite of late-Universe distance probes---including the completed SDSS-IV BAO consensus compilation, DESI DR2 BAO, transverse/angular BAO (BAOtr), and the Cepheid-calibrated PantheonPlus SN~Ia likelihood (PP\&SH0ES)---we have tested both the resulting late-time dynamics and the internal consistency of current low-redshift datasets.

Our principal findings highlight a strong dataset dependence of the reconstructed late-time expansion history within the $w_0w_a$CDM (CPL) parametrization. When constrained by CMB data alone, CPL admits an extended phantom-like region of parameter space and can drive the inferred present-day deceleration parameter into $q_0<-1$ (super-acceleration, $\dot H>0$). However, as discussed in Sec.~III\,B, CMB-only constraints primarily fix the distance to last scattering and allow a broad geometric degeneracy in $(H_0,\Omega_{\rm m},w_0,w_a)$; in this situation the $q_0<-1$ region arises in the degeneracy/extrapolation tail of the CPL posterior in the absence of low-redshift distance anchors, rather than constituting a robust inference about the true late-time expansion state. The inclusion of high-precision DESI DR2 BAO pulls the CPL reconstruction in the opposite direction, favoring a weakly accelerating or nearly coasting present-day Universe ($q_0\simeq 0$), whereas combining CMB with PP\&SH0ES and BAOtr yields a more conventional, moderately accelerating expansion ($-1<q_0\lesssim 0$) and substantially reduces the Hubble tension. We also find that the combinations that tightly constrain the CPL evolution parameter (notably SDSS/DESI and the PP\&SH0ES-inclusive fits) favor $w_a<0$, and the inferred high-redshift asymptote $w(z\to\infty)=w_0+w_a$ is phantom-like ($< -1$) across all dataset combinations considered (see~\cref{tab:parameters}); nevertheless, $q(z)$ increases and approaches the $\Lambda$CDM-like matter-dominated limit $q\to 1/2$ by $z\gtrsim 0.5$ (i.e.\ within the post-recombination regime relevant to the late-time probes considered here), indicating that the total expansion is non-phantom at high redshift because dark energy is already subdominant there. Taken together, these results show that CPL reconstructions are not dataset-stable when mutually pulling low-redshift distance information is included, underscoring the limited adequacy of the simple two-parameter CPL form as a universal phenomenological description of late-time cosmic expansion.

A critical result of our analysis is the identification of a non-negligible mismatch among low-redshift BAO distance information. In particular, BAOtr and the standard three-dimensional BAO determinations from DESI/SDSS prefer different distance ratios at $z\lesssim0.5$, which propagates into markedly different reconstructions of the background expansion history $H(z)$ and the deceleration parameter $q(z)$ within a flexible late-time ansatz such as CPL. At the observable level, the common $z=0.510$ transverse comparison gives $T_\theta=3.34\sigma$ for BAOtr versus DESI LRG1 and $T_\theta=2.87\sigma$ for BAOtr versus SDSS DR12, while the local BAOtr-interpolated SDSS comparison gives $\chi^2_{\rm diff}=12.35$ for two bins, corresponding to $3.08\sigma$. These values should be read with the covariance caveats discussed in Sec.~\ref{subsecIIID}, but they show that the BAOtr--3D BAO mismatch is already visible at the level of the low-redshift BAO observables, not only after projection into CPL parameter space. The same tension is visible in the inferred $(w_0,w_a)$ parameter space: the regions preferred by CMB+DESI and by combinations involving PP\&SH0ES(+BAOtr) do not fully overlap and are inconsistent at the $\gtrsim2\sigma$ level, even though both independently disfavor the $\Lambda$CDM point $(w_0,w_a)=(-1,0)$ within CPL.

Model-comparison statistics further illuminate this contingent picture. While CPL (with two additional parameters) can reduce the best-fit $\chi^2_{\min}$ across all dataset combinations, the Bayesian evidence $\Delta\ln\mathcal{Z}$---which accounts for the enlarged CPL parameter volume---is strongly dataset-dependent. In particular, CPL is strongly favored over $\Lambda$CDM primarily for combinations including PP\&SH0ES and/or BAOtr, whereas CMB+DESI yields only inconclusive evidence for CPL and the inferred $H_0$ remains in strong tension with local distance-ladder determinations. This pattern suggests that the apparent CPL ``alleviation'' of the Hubble tension is not universal, but depends sensitively on which specific late-Universe distance information is included.

Within CPL, the near stability of $r_{\rm d}$ is expected because the pre-recombination sector is fixed. It is still useful diagnostically: the dataset-dependent shifts in $H_0$ are absorbed mainly by $r_{\rm d}H_0$ and the reconstructed late-time history $E(z)$, rather than by a fitted change in $r_{\rm d}$. This is not a direct test or exclusion of early-dark-energy models, which must be analyzed in their own parameter spaces. The possible indirect implication is conditional: if PP\&SH0ES and BAOtr are reliable low-redshift anchors, early-time-only solutions that retain a $\Lambda$CDM-like late-time expansion may be insufficient. For DESI/SDSS combinations, however, $\Lambda$CDM remains viable or even preferred.

Our analysis reveals limitations of the simple two-parameter CPL parametrization as a universal phenomenological description of late-time expansion. While CPL can fit some low-redshift dataset combinations simultaneously (e.g. CMB+PP\&SH0ES+BAOtr), its reconstructed expansion history depends strongly on which BAO dataset is included: adding the high-precision DESI DR2 BAO measurements pulls the fit toward a different region of the CPL parameter space than that preferred by BAOtr/PP\&SH0ES, so there is no single CPL reconstruction that remains consistent across all current low-redshift distance probes. The resulting spread in reconstructed $H(z)$ and $q(z)$ across dataset combinations---including the appearance of a $q_0<-1$ tail in the CMB-only CPL case---should be interpreted as a manifestation of CPL degeneracy/extrapolation in the absence of late-time distance anchors, rather than as a robust inference that the real Universe requires super-acceleration. Future progress will therefore likely require a dual-path approach: exploring more flexible and/or physically motivated dark-energy or modified-gravity descriptions with richer late-time dynamics, alongside continued scrutiny of potential residual systematics and cross-calibration among low-redshift probes.

Finally, the phantom-like CPL asymptote $w(z\to\infty)=w_0+w_a< -1$ should be viewed as an extrapolated descriptor of low-redshift posterior correlations, not as a direct early-dark-energy measurement. In the standard-fluid interpretation of CPL, the corresponding decrease of $\rho_{\rm de}(z)$ can only asymptote toward $\rho_{\rm de}\to0$ at high redshift. This behavior motivates, but does not demonstrate, sign-changing scenarios in which $\rho_{\rm de}$ crosses zero and becomes negative at sufficiently high redshift, such as braneworld~\cite{Sahni:2002dx,Mishra:2025goj}, late-time AdS-to-dS(-like) transition~\cite{DiValentino:2017rcr,Akarsu:2019hmw,Akarsu:2021fol,Akarsu:2022typ,Akarsu:2023mfb,Anchordoqui:2023woo,Anchordoqui:2024gfa,Akarsu:2024qsi,Akarsu:2024eoo,Gokcen:2026pkq}, and $f(T)$ teleparallel gravity~\cite{Akarsu:2024nas} models. This broader interpretation is also consistent with recent late-time, model-agnostic comparisons showing that smooth low-dimensional EoS parametrizations, including CPL, can reproduce the directly constrained expansion history while compressing localized intermediate-redshift structure in derivative-sensitive quantities such as $q(z)$ and in the GR-mapped effective dark-energy sector, particularly around $z\sim1.5$--$2$~\cite{Akarsu:2026anp,Akarsu:2026pom}. Testing whether such physically distinct extensions give a more dataset-stable account of the low-redshift distance data is left for future work.

\begin{acknowledgments}
This work was supported by the National Key Research and Development Program of China (Nos. 2022YFA1602903 and 2023YFB3002501), the National Natural Science Foundation of China (Nos. 12588202 and 12473002), and the China Manned Space Program (Grant No. CMS-CSST-2025-A03). S.K. gratefully acknowledges the support of Startup Research Grant from Plaksha University (File No.\ OOR/PU-SRG/2023-24/08). A.J.S.C.\ acknowledges Conselho Nacional de Desenvolvimento Cient\'{\i}fico e Tecnol\'ogico (CNPq; National Council for Scientific and Technological Development) for partial financial support (Grant No.~305881/2022-1) and Funda\c{c}\~ao da Universidade Federal do Paran\'a (FUNPAR; Paran\'a Federal University Foundation) through public notice 04/2023-Pesquisa/PRPPG/UFPR for partial financial support (Process No.~23075.019406/2023-92), as well as the financial support of the NAPI ``Fen\^omenos Extremos do Universo'' of Funda\c{c}\~ao de Apoio \`a Ci\^encia, Tecnologia e Inova\c{c}\~ao do Paran\'a (NAPI F\'ISICA--FASE~2), under protocol No.~22.687.035-0. \"{O}.A.\ acknowledges the support by the Turkish Academy of Sciences in scheme of the Outstanding Young Scientist Award (T\"{U}BA-GEB\.{I}P). This article is based upon work from the COST Action CA21136 ``Addressing observational tensions in cosmology with systematics and fundamental physics'' (CosmoVerse), supported by COST (European Cooperation in Science and Technology).
\end{acknowledgments}

\appendix

\section{BAO data compilation}
\label{app:bao_data}

This appendix lists the BAO measurements used in our likelihoods. The tables report central values and diagonal $1\sigma$ uncertainties; correlated DESI DR2 and SDSS/eBOSS points are evaluated using full covariance matrices or public likelihood grids.

\begin{table*}[t!]
\centering
\caption{\label{tab:baotr_data}
Angular/transverse BAO measurements used in the BAOtr likelihood~\cite{Nunes:2020hzy}. The measured quantity is $\theta_{\rm BAO}=r_{\rm d}/D_{\rm M}$ in degrees. The last two columns give the equivalent transverse distance ratio $D_{\rm M}/r_{\rm d}$ and its symmetric first-order propagated $1\sigma$ uncertainty, obtained by converting $\theta_{\rm BAO}$ to radians. In Fig.~\ref{fig:d_z}, the BAOtr points are displayed with asymmetric error bars under the inverse transformation from $\theta_{\rm BAO}$ to $D_{\rm M}/r_{\rm d}$, following the plotting convention used for the original figure.}
\setlength{\tabcolsep}{6pt}
\renewcommand{\arraystretch}{1.06}
\footnotesize
\begin{tabular}{ccccc l}
\toprule
$z$ & $\theta_{\rm BAO}$ [deg] & $\sigma_\theta$ [deg] & $D_{\rm M}/r_{\rm d}$ & $\sigma(D_{\rm M}/r_{\rm d})$ & Source sample \\
\midrule
0.110 & 19.80 & 3.26 & 2.894 & 0.476 & SDSS DR7 blue galaxies \\
0.235 & 9.06 & 0.23 & 6.324 & 0.161 & SDSS DR10/DR11 photometric LRGs \\
0.365 & 6.33 & 0.22 & 9.051 & 0.315 & SDSS DR10/DR11 photometric LRGs \\
0.450 & 4.77 & 0.17 & 12.012 & 0.428 & SDSS DR10 LRGs \\
0.470 & 5.02 & 0.25 & 11.414 & 0.568 & SDSS DR10 LRGs \\
0.490 & 4.99 & 0.21 & 11.482 & 0.483 & SDSS DR10 LRGs \\
0.510 & 4.81 & 0.17 & 11.912 & 0.421 & SDSS DR10 LRGs \\
0.530 & 4.29 & 0.30 & 13.356 & 0.934 & SDSS DR10 LRGs \\
0.550 & 4.25 & 0.25 & 13.481 & 0.793 & SDSS DR10 LRGs \\
0.570 & 4.59 & 0.36 & 12.483 & 0.979 & SDSS DR11 LRGs \\
0.590 & 4.39 & 0.33 & 13.051 & 0.981 & SDSS DR11 LRGs \\
0.610 & 3.85 & 0.31 & 14.882 & 1.198 & SDSS DR11 LRGs \\
0.630 & 3.90 & 0.43 & 14.691 & 1.620 & SDSS DR11 LRGs \\
0.650 & 3.55 & 0.16 & 16.140 & 0.727 & SDSS DR11 LRGs \\
2.225 & 1.77 & 0.31 & 32.370 & 5.669 & SDSS DR12 photometric quasars \\
\bottomrule
\end{tabular}
\end{table*}

\begin{table*}[t!]
\centering
\caption{\label{tab:desi_data}
DESI DR2 BAO measurements used in this work~\cite{DESI:2025zgx}. The quoted uncertainties are the square roots of the diagonal entries of the DESI covariance matrix. For anisotropic tracers, the full covariance between the paired $D_{\rm M}/r_{\rm d}$ and $D_{\rm H}/r_{\rm d}$ measurements is used in the likelihood; the correlation coefficient $\rho_{M,H}$ is shown for reference.}
\setlength{\tabcolsep}{5pt}
\renewcommand{\arraystretch}{1.06}
\footnotesize
\begin{tabular}{llccc}
\toprule
Tracer & $z_{\rm eff}$ & Observable & Value $\pm1\sigma$ & $\rho_{M,H}$ \\
\midrule
BGS & 0.295 & $D_{\rm V}/r_{\rm d}$ & $7.942\pm0.076$ & -- \\
LRG1 & 0.510 & $D_{\rm M}/r_{\rm d}$ & $13.588\pm0.168$ & \multirow{2}{*}{$-0.452$} \\
LRG1 & 0.510 & $D_{\rm H}/r_{\rm d}$ & $21.863\pm0.429$ & \\
LRG2 & 0.706 & $D_{\rm M}/r_{\rm d}$ & $17.351\pm0.180$ & \multirow{2}{*}{$-0.395$} \\
LRG2 & 0.706 & $D_{\rm H}/r_{\rm d}$ & $19.455\pm0.334$ & \\
LRG3+ELG1 & 0.934 & $D_{\rm M}/r_{\rm d}$ & $21.576\pm0.162$ & \multirow{2}{*}{$-0.347$} \\
LRG3+ELG1 & 0.934 & $D_{\rm H}/r_{\rm d}$ & $17.641\pm0.201$ & \\
ELG2 & 1.321 & $D_{\rm M}/r_{\rm d}$ & $27.601\pm0.325$ & \multirow{2}{*}{$-0.398$} \\
ELG2 & 1.321 & $D_{\rm H}/r_{\rm d}$ & $14.176\pm0.225$ & \\
QSO & 1.484 & $D_{\rm M}/r_{\rm d}$ & $30.512\pm0.764$ & \multirow{2}{*}{$-0.494$} \\
QSO & 1.484 & $D_{\rm H}/r_{\rm d}$ & $12.817\pm0.518$ & \\
Ly$\alpha$ & 2.330 & $D_{\rm H}/r_{\rm d}$ & $8.632\pm0.101$ & \multirow{2}{*}{$-0.431$} \\
Ly$\alpha$ & 2.330 & $D_{\rm M}/r_{\rm d}$ & $38.989\pm0.532$ & \\
\bottomrule
\end{tabular}
\end{table*}

\begin{table*}[t!]
\centering
\caption{\label{tab:sdss_data}
Completed SDSS-IV/eBOSS BAO measurements used as the pre-DESI BAO compilation~\cite{eBOSS:2020yzd}. The table lists the central values and representative $1\sigma$ uncertainties of the observables entering the likelihoods. Gaussian measurements are evaluated with their covariance matrices; the MGS, ELG, and Ly$\alpha$ entries are implemented through the public one- or two-dimensional likelihood tables.}
\setlength{\tabcolsep}{5pt}
\renewcommand{\arraystretch}{1.06}
\footnotesize
\begin{tabular}{llcc}
\toprule
Sample & $z_{\rm eff}$ & Observable & Value $\pm1\sigma$ \\
\midrule
MGS & 0.150 & $D_{\rm V}/r_{\rm d}$ & $4.466\pm0.168$ \\
BOSS DR12 galaxies & 0.380 & $D_{\rm M}/r_{\rm d}$ & $10.234\pm0.169$ \\
BOSS DR12 galaxies & 0.380 & $D_{\rm H}/r_{\rm d}$ & $24.981\pm0.729$ \\
BOSS DR12 galaxies & 0.510 & $D_{\rm M}/r_{\rm d}$ & $13.366\pm0.204$ \\
BOSS DR12 galaxies & 0.510 & $D_{\rm H}/r_{\rm d}$ & $22.317\pm0.572$ \\
eBOSS LRG & 0.698 & $D_{\rm M}/r_{\rm d}$ & $17.858\pm0.328$ \\
eBOSS LRG & 0.698 & $D_{\rm H}/r_{\rm d}$ & $19.326\pm0.533$ \\
eBOSS ELG & 0.845 & $D_{\rm V}/r_{\rm d}$ & $18.33\pm0.62$ \\
eBOSS QSO & 1.480 & $D_{\rm M}/r_{\rm d}$ & $30.688\pm0.798$ \\
eBOSS QSO & 1.480 & $D_{\rm H}/r_{\rm d}$ & $13.261\pm0.552$ \\
Ly$\alpha$ auto & 2.334 & $D_{\rm M}/r_{\rm d}$ & $37.60\pm1.90$ \\
Ly$\alpha$ auto & 2.334 & $D_{\rm H}/r_{\rm d}$ & $8.93\pm0.28$ \\
Ly$\alpha\times$QSO & 2.334 & $D_{\rm M}/r_{\rm d}$ & $37.30\pm1.70$ \\
Ly$\alpha\times$QSO & 2.334 & $D_{\rm H}/r_{\rm d}$ & $9.08\pm0.34$ \\
\bottomrule
\end{tabular}
\end{table*}

\clearpage

\bibliographystyle{apsrev4-2_mod_yearfix}
\bibliography{bibfile}

\end{document}